\documentclass[a4paper,fleqn, 10pt]{cas-sc}

\usepackage[numbers]{natbib}

\usepackage{hyperref}

\usepackage{amsthm}
\usepackage{float}
\usepackage{caption}
\usepackage{subcaption}
\usepackage{tikz}

\usepackage{soul}

\usetikzlibrary{arrows}
\def\tsc#1{\csdef{#1}{\textsc{\lowercase{#1}}\xspace}}
\tsc{WGM}
\tsc{QE}
\tsc{EP}
\tsc{PMS}
\tsc{BEC}
\tsc{DE}

\begin{document}
\let\WriteBookmarks\relax
\def\floatpagepagefraction{1}
\def\textpagefraction{.001}
\shorttitle{Fracture in disordered solids}
\shortauthors{Hudson Borja da Rocha et~al.}

\title [mode = title]{Mean field fracture in disordered solids: statistics of fluctuations}                      
%
%
%

\author[1]{Hudson Borja da Rocha\corref{correspondingauthor}}[]
\cormark[1]

\ead{hudson.borja-da-rocha@polytechnique.edu}

\author[2]{Lev Truskinovsky}[]

\ead{lev.truskinovsky@espci.fr}


\address[1]{CIRB, CNRS UMR 7241, INSERM U1050, Collège de France and PSL Research University, F-75005 Paris, France}

\address[2]{PMMH, CNRS-UMR 7636 PSL-ESPCI, 10 Rue Vauquelin, 75005 Paris, France}

\cortext[cor1]{Corresponding author. \\ \textit{E-mail addresses:} \href{mailto:hudson.borja-da-rocha@polytechnique.edu}{hudson.borja-da-rocha@polytechnique.edu} (H. Borja da Rocha), \href{mailto:lev.truskinovsky@espci.fr}{lev.truskinovsky@espci.fr} (L. Truskinovsky) }
%

\begin{abstract}
Power law distributed fluctuations are known to accompany \emph{terminal} failure in disordered brittle solids. The associated intermittent scale-free behavior is of interest from the fundamental point of view as it emerges universally from an intricate interplay of threshold-type nonlinearity,  quenched disorder, and long-range interactions. We use the simplest mean-field description of such systems to show that they can be expected to undergo a transition between brittle and quasi-brittle (ductile) responses. While the former is characterized by a power law distribution of avalanches, in the latter,  the statistics of avalanches is predominantly Gaussian. The realization of a particular regime depends on the variance of disorder and the effective rigidity represented by a combination of elastic moduli.  We argue that the robust criticality, as in the cases of earthquakes and collapsing porous materials,  indicates the self-tuning of the system towards the boundary separating brittle and ductile regimes. 
\end{abstract}

%


\maketitle


\section{Introduction}

While many engineering  aspects of  fracture in intrinsically disordered solids have been thoroughly investigated~\citep{krajcinovic2000damage,bazant2017probabilistic,curtin1998stochastic}, the statistical  properties of fracture-induced fluctuations  have become a subject of focused  research only recently~\citep{herrmann2014statistical,Alava_2006, Fortin2009, Petri_PRL_1994, Tordesillas_FM_2020}. The interest towards fluctuations is mostly driven by experimental observations revealing that acoustic emission  measured prior to \emph{terminal}  fracture and the roughness of the ensuing fracture surface  are  described by scale-free laws, indicating turbulence-type complexity and suggesting criticality~\citep{BONAMY20111,Bouchbinder_2014,Girard_2010, Vives_PRL_2018, Weiss_PRL_2019}. Power law distributed spatial and temporal fluctuations in disordered brittle solids are  also of interest from the fundamental point of view because of the intricate interplay in the underlying  processes between threshold-type, strong nonlinearity, quenched disorder and long-range interactions~\citep{biswas2015statistical,SAHIMI1998213}. Of practical importance  is the accurate prediction of  the  terminal failure  through the  identification of the  precursors for the catastrophic   coalescence  of micro-fractures~\citep{Bazant_2019,Purusattam_2019, Sornette_PNAS_2002}.

The observed scale-free behaviors in fracturing solids have been linked, by some authors, to thermodynamic spinodal points associated with first-order phase transitions \citep{Alava_2006, Zapperi_PRL_1997} and, by other authors, to critical points or second-order phase transitions \citep{Moreno_PRL_2000,  Weiss_PNAS_2014, Weiss_PRL_2019, Ciliberto_PRL_1997}. An additional complication here is that athermal fracture can be modeled in two different ways: as an incremental \emph{global} energy minimization (GM) phenomenon (zero-temperature limit of a thermal equilibrium response), e.g.  \citep{FRANCFORT19981319, bourdin2008variational} or as an incremental \emph{local} energy minimization keeping the system in the state of marginal stability (MS) (zero viscosity limit of an overdamped response), e.g. \citep{selinger_1991, Selinger_PRB_1991, PT_JMPS_2005}.

The statistical physics approach to fracture is usually based on direct simulation of discrete models, which carry a regularizing internal length scale, allow for a relatively simple introduction of disorder  \citep{Hansen_Roux_2000} and are suitable for the inclusion of thermal noise \citep{Politi_2002, Roux_PRE_2000, BRT_AAM_2019}. In such models, the continuum medium is represented by a network of elastic elements while the disorder is modeled either by random failure thresholds or by random elimination of the elements. Crucially, the discrete models retain the long-range nature and the tensorial structure of the elastic interactions \citep{Nikolic2018,ostoja2007microstructural}. 

Many important discoveries in statistical fracture mechanics were made based on the study of the discrete random fuse model (RFM) in which a scalar field represents displacements, elasticity is linear, but the threshold nature of fracture is preserved  \citep{deArcangelis_1985}. And still, RFM is too complex to be amenable to analytical study, so further simplifications were made to allow for theoretical analysis. The desired analytical transparency was achieved (without losing the main effects) in the framework of the exactly solvable global load sharing (GLS) fiber bundle model (FBM) \citep{hansen2015fiber}, which can be seen as a mean-field approximation of the RFM.  More comprehensive continuum modeling approaches, including phase field~\citep{Berthier_JMPS_2017, berthier2021damage, Gorgogianni_JAM_2020} and mesoscopic damage models~\citep{Girard_2010, Patinet2014} are still mostly numerical.

In this paper, we use the augmented GLS model to show that both spinodal and critical \emph{scaling behaviors} are relevant near the threshold of the brittle-to-ductile transition, which is characteristic for such systems \citep{Liu_JPSB_2019, SELEZNEVA201820}. The ductile response is understood here in the sense of stable development of small avalanches representing micro-failure events \citep{Christensen_2018}, while the brittle response is defined by an abrupt macro-failure event representing system-size instability \citep{Papanikolaou2019, Berthier_PRM_2019}. This abruptness is the consequence of stress concentration in the system that induces an autocatalytic bond-breaking process. In ductile systems such catastrophic breaking is absent, and as a consequence, the mechanical response in the ductile regime is gradual, persisting after the stress peak. The signature of ductile fracture is that the fracture toughness is large, the inelastic deformation is spatially diffuse, and the failure is gradual.  Instead, the fracture toughness associated with a brittle fracture is small, and the inelastic deformation is localized. Understanding the microscopic spatio-temporal processes behind the stochastization near the brittle-to-ductile  (BTD) transition and revealing the statistical structure of the associated intermittent fluctuations is one of the most important challenges in statistical fracture mechanics.

To study these intriguing phenomena \emph{analytically} we needed to go beyond the simplest GLS model, even though it has proved successful in describing a variety of physical phenomena from the failure of textiles and acoustic emission in loaded composites to earthquake dynamics \citep{NECHAD20051099, XIA2001273, Pradhan:2010aa}. The problem is that in all these applications, the fracture development could be studied in a stress control (soft device) setting, where the ultimate failure is necessarily \emph{brittle}.  As a result, the terminal fracture is accompanied by standard fluctuations exhibiting universal scaling of spinodal type \citep{hansen,hansen2015fiber, Sornette, Kloster}.  Various "non-democratic" settings of FBM, implying local load sharing, have also been studied in the soft device setting, which ultimately obscures the BDT transition \citep{Pradhan:2010aa, Patinet2014, Roux_IJSS_1999}.


To capture the ductile behavior and to be able to \emph{tune} the system to criticality, we had to address failure under strain control (hard device) \citep{Roux_IJSS_1999}. To this end, in this paper, we introduced two seemingly innocent extensions of the standard FBM in the form of additional internal (series to fibers) and external (series to the bundle) linear springs. The main advantage of the new framework is that by changing the relative  stiffness
 of the internal and external springs, one can simulate the crossover from brittle to ductile response. Interestingly,  we found that the augmented FBM exhibits different types of scaling in these two regimes: brittle failure emerges as a supercritical, while ductile failure comes out as a subcritical phenomenon. The \emph{critical} behavior can be associated with the BDT transition only, and we show that due to the superuniversality of mean-field models \citep{PhysRevB.89.104201}, the MS and GM critical exponents are the same. 

The augmented  FBM also reveals the parallel roles of disorder and  the effective rigidity (represented here  by the strategic ratio of elastic moduli) as the two main \emph{regulators} of the BDT transition \citep{Merkel6560, Vitelli_PNAS_2016, Xiaoming_PRM_2017, Dussi_PRL_2020, BRT_PRL_2020,richard2021brittle,david2021finitesize}  separating ductile behavior at low  rigidity and high disorder from the brittle behavior at low disorder and high  rigidity. Since both ductility and brittleness represent the material response outside the elastic limit,  the emergence of an elastic control through the effective rigidity  may appear counter-intuitive in this context. We recall, however,   that the inelastic mechanical response is intimately related to the microscopic structure of the crystal.  For instance,  the manner in which  stress is redistributed in response to intrinsic instabilities (representing inelastic effects) depends on both the strength and the topology of atomic bonding. In particular, the parameters of the BDT transition  are  known to be highly sensitive to the internal connectivity of the crystal lattice represented  either by the  Poisson's ratio or by the ratio of bulk to shear modulus \citep{Greaves2011}.  In this paper, we show that in addition to influencing the failure mode, such ratios can also fundamentally affect the  fluctuational precursors of the macro-instability, which can be then   used  to    build  statistical  predictors of the ultimate structural failure.

The rest of the paper is organized as follows. In Section 2, we introduce the model and analyze the averaged response. The empirical avalanche distributions, emerging from direct numerical simulation of the model, are studied in Section 3. Then in Section 4, we build a mapping between the micro-fracture avalanches and biased random walks which allows us to compute analytically their statistical characteristics in both MS and GM regimes. The asymptotic scaling laws for brittle, critical, and ductile behaviors are derived in Section 5.  The finite-size scaling is studied in Section 6. Finally, in Section 7, we present our conclusions.

Some of the results of this paper were first announced in~\citep{BRT_PRL_2020}.


\section{The model}

\begin{figure}[ht]
\centering
\includegraphics[width = 0.9\textwidth]{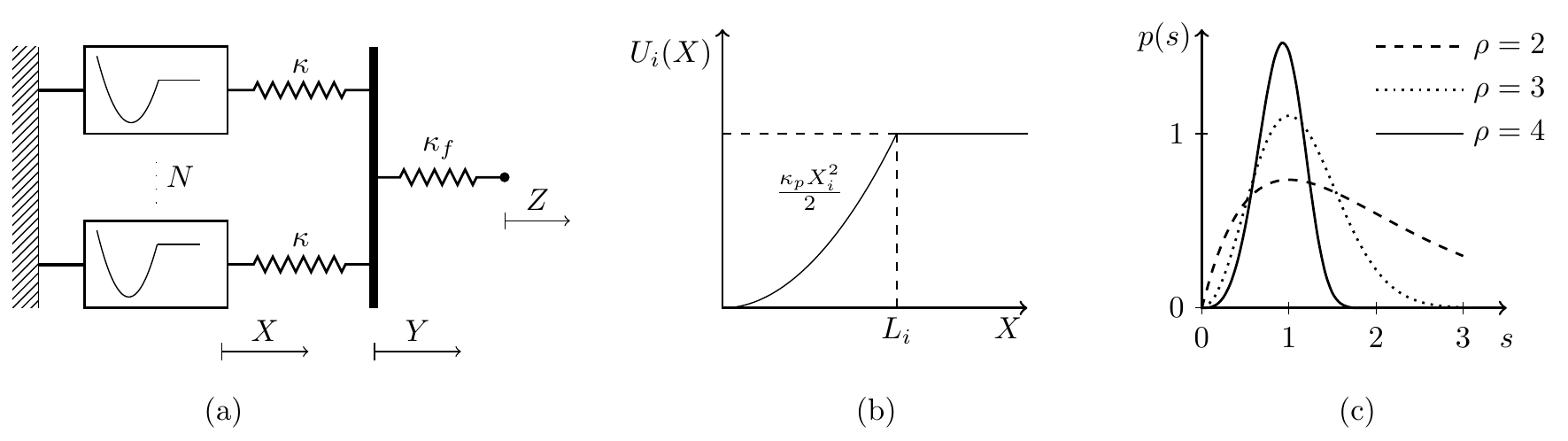}
\caption{(a) Mechanical model of the system, (b)  the potential energy of a single bond and (c)   Weibull distribution of the debonding thresholds $p(s)$ at different values of the parameter $\rho$. }
\label{fig:mecmod}
\end{figure}

Consider $N$ parallel breakable elastic links  and suppose that a link  with index $i=1,\dots,N$ is characterized by a  breaking threshold $L_i$. Assume further that the system is loaded through a load redistributing medium modeled as  an external spring with  stiffness
 $\kappa_f$ \citep{Roux_IJSS_1999}. The total energy of the system can be written as,
\begin{equation}\label{eq:pot21}
\displaystyle E =\sum_{i}^N U_i(X_i)+\frac{\kappa}{2}\sum_{i}^N (Y-X_i)^{2}+\frac{\kappa_f}{2}(Z-Y)^2,
\end{equation}
where $Y$ is a position of the connecting backbone and $Z$ is the total elongation serving as  the controlling parameter, see  Fig.~\ref{fig:mecmod}(a). The nonlinear breakable element is characterized by  the  potential,
$
U_i(X)=  \kappa_{p}X^{2}/2$  for $X\leq L_i$ and 
$
U_i(X)=\kappa_{p}L_{i}^{2}/2$ for $ X>L_i.$
The variable $X_i$ represents the internal degree of freedom of the unit  $i$. For values of $X_i$ smaller than $L_i$ the element is attached to the backbone and bears the same force as the other bound units. When the variable $X_i$ reaches the threshold $L_i$ the $i$'th element breaks, dissipating (or transforming into a non mechanical form) the energy $\kappa_{p}L_{i}^{2}/2$. 

It will be convenient to work with dimensionless variables. We define the reference scale $\ell_0$ to be the resting length of the linear springs, and introduce  dimensionless variables $x_i=X_i/\ell_0$, $y=Y/\ell_0$, $z=Z/\ell_0$ and $l_i=L_i/\ell_0$.  The remaining non-dimensional parameters of the problem are 
$
\displaystyle{\lambda= \kappa/\kappa_p}
$ and  $
\displaystyle{\lambda_f= \kappa_f/(N \kappa_p)}.
$
The dimensionless energy per element in a hard device
 is given by
\begin{equation}\label{eq:ehd}
\mathcal{H}(\boldsymbol{x},y,z)={\displaystyle \frac{1}{N}\sum_{i=1}^{N}}\left[u_i(x_{i})+\frac{\lambda}{2}\left(y-x_{i}\right)^{2}\right]+\frac{\lambda_{f}}{2}(z-y)^{2}.
\end{equation}
where,
$ u_i(x)= x^{2}/2$  for $x\leq l_i$ and $u_i(x)= l_{i}^{2}/2$ for $x>l_i.$
We assume that the failure thresholds $l_i$ are \emph{disordered} and represented by independent random variables each described by  the same probability density $p(s)$, and the same cumulative distribution  $P(s)=\int_{0}^{s}p(y)dy$. 
The soft device loading can be seen as a limiting case of the hard device loading; it can be obtained if we assume that the outer spring is infinitely soft $\lambda_f\rightarrow 0$. In this limit  $z\rightarrow \infty$, but if simultaneously $\lambda_f z \rightarrow f$ we obtain the system loaded by the force $f$.

To explore the set of the metastable states we need to first equilibrate the system with respect to the internal variables $x_ i$'s and $y$. To this end, we need to  solve the system of equations:
\begin{equation}\label{eq:eqhd}
\begin{cases}
\begin{array}{cl}
{\displaystyle \frac{{\partial \mathcal{H}}}{\partial x_{i}}}&=0,\mbox{for all} \,1\leq i\leq N \\[10pt]
{\displaystyle \frac{\partial \mathcal{H}}{\partial y}}&=0.
\end{array}\end{cases}
\end{equation}
Equilibration in $\boldsymbol{x}$ gives $u'(x_i)=\lambda(y-x_{i})$, or more explicitly, $x_i=x_0$ for $ x_i\leq l_i$ and $x_i=x_{00}$ for $ x_i >l_i$, where 
$x_0= \lambda y/(\lambda+1)$ and $x_{00}=y$.
The equilibration in $y$ gives,
\begin{equation}\label{eq:equilibrium_y}
y(\boldsymbol{x},z)={\displaystyle \frac{1}{\lambda+\lambda_{f}}{\displaystyle \left(\lambda_{f}z+\lambda\frac{1}{N}\sum_{i=1}^{N}x_{i}\right)}}.
\end{equation}
We see the mean-field nature of the coupling through Eq.~\eqref{eq:equilibrium_y}: the variable $y$ is affected by the average value of $x_i$. Using permutational invariance we can write 
\begin{equation}
{\displaystyle \frac{1}{N}\sum_{i=1}^{N}x_{i}=\frac{k}{N} y+\frac{N-k}{N}\frac{\lambda y}{\lambda+1}}
\end{equation}
where $k$ is the number of broken bonds. This representation allows us to write the equilibrium  elongation $\hat{y}$ as a function of the   total elongation $z$:
\begin{equation}\label{eq:equiy}
\hat{y}(k,z)={\displaystyle \frac{\Lambda z}{ 1-k/N +\Lambda }}.
\end{equation}
Here we first encounter one of our main nondimensional parameters
\begin{equation}\label{eq:Lambda}
\Lambda= \lambda_f/\lambda_*.
\end{equation}
where $\lambda_*= \lambda/(\lambda+1)$  is the   dimensionless effective stiffness of  an individual   breakable  link.  The synthetic parameter $\Lambda$  can be perceived as inversely correlated with the system's rigidity. It can be viewed as a  measure of the overall  'malleability' of the structure.

The equilibrium  values of $x_i$ for  the closed  and open  configurations are, respectively, 
\begin{equation}\label{eq:x0}
\hat{x}_{0}(k,z)={\displaystyle \frac{\lambda_{f} z}{1-k/N +\Lambda}},\, \hat{x}_{00}(k,z)={\displaystyle \frac{ \Lambda z}{ 1 - k/N +\Lambda}}.
\end{equation}
Each value of $k$ defines an equilibrium branch extending between the limits $(z_{inf}(k),\, z_{sup}(k))$. Due to the presence of  the  backbone, individual  elements   necessarily fail in sequence according to the value of their thresholds. Let $\bar{x}_i, i=1,...,N$ be the ordered sequence of failure thresholds $l_i$: $\bar{x}_1\leq \bar{x}_2\leq \dots\leq \bar{x}_N $.   In terms of the parameters  $\bar{x}_i $ we can formulate  the constraints as
  $\hat{x}_{0}(k,z)<\bar{x}_k$   and $\hat{x}_{00}(k,z)> \bar{x}_k$. Therefore  $\Lambda z_{inf} =\lambda_f z_{sup}=[1 - k/N  + \Lambda ] \bar{x}_k$. 
   The special cases are the homogeneous configurations with $k=0$, defined for $z\in (-\infty,z_{sup}(k=0)]$ and with $k=N$, defined for $z\in[z_{inf}(k=N),\infty)$. 
%

To analyze the stability of the obtained equilibrium states,  we   compute the Hessian of the energy $\tilde{\mathcal{H}}=N \mathcal{H}(\boldsymbol{x},y)$. We obtain 
\begin{equation}
\underline{\underline{\boldsymbol{ \mathcal{M}}}}=
\begin{pmatrix}
\tilde H_1 & 0 & \dots & 0 & -\lambda \\ 
0 & \ddots & \ddots & \vdots & \vdots \\ 
\vdots & \ddots & \ddots & 0 & \vdots \\ 
0 & \dots & 0 & \tilde H_N & -\lambda \\ 
-\lambda & \dots & \dots & -\lambda & N(\lambda+\lambda_f)
\end{pmatrix} ,
\end{equation}
where $\tilde H_i=\lambda+1 $ for $1\leq i < N-k$  and  $\tilde H_i=\lambda$ for $N-k \leq i \leq N$.
The first $N$  principal minors of  $\mathcal{M}$ are   the product of diagonal therms $\tilde H_i$ and are therefore always positive. The last principal minor is
\begin{equation}
\det(\mathcal{\underline{\underline{\boldsymbol{ \mathcal{M}}}}})=\prod_{i=1}^N \tilde H_i \sum_{i=1}^{N} \left(\lambda+\lambda_f- \frac{\lambda^2}{\tilde H_i}\right) \geq 0.
\end{equation}
 Therefore,   all  equilibrium configurations obtained above are stable. The unstable configurations must contain at least one element in the 'spinodal' state    $x_i=l_i$.

\begin{figure}[ht]
\centering
\begin{subfigure}[b]{0.49\textwidth}
\centering
\includegraphics[scale=1]{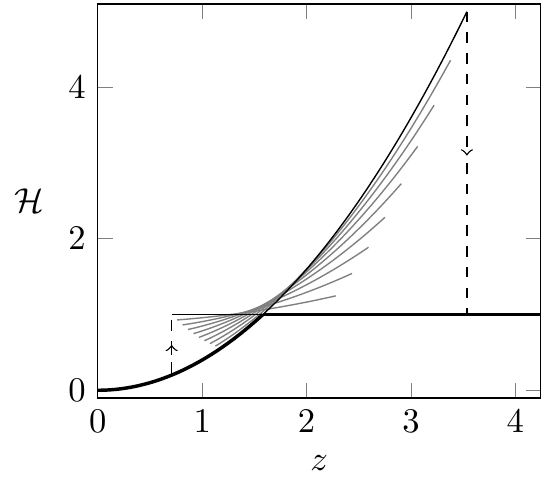}
\caption{}\label{fig:eneHD}
\end{subfigure}
\begin{subfigure}[b]{0.49\textwidth}
\centering
\includegraphics[scale=1]{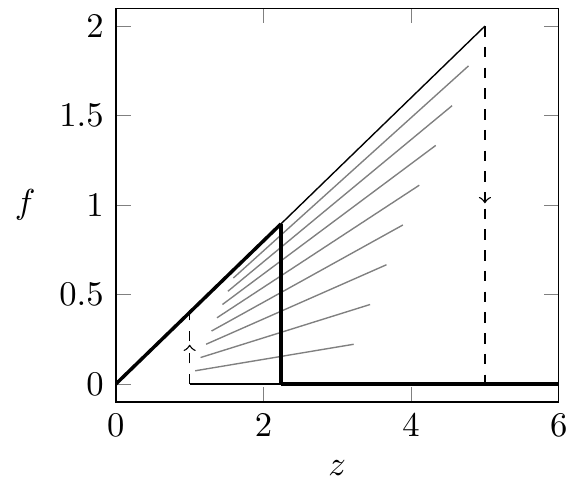}
\caption{}
\end{subfigure}
\caption{Mechanical response of the model in a hard device. Solid black lines, global minimum, gray lines metastable states. Parameters are $N=10$, $\lambda=1$, $\lambda_f=1$ and $l=\sqrt{2}$.}\label{fig:HD}
\end{figure}

\begin{figure}[ht]
\begin{subfigure}[b]{0.49\textwidth}
\centering
\includegraphics[scale=1]{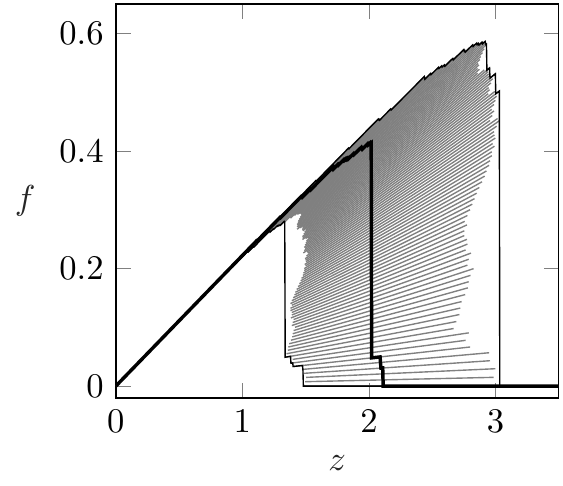}
\caption{}
\label{fig:Tension_rho_4_lf_04}
\end{subfigure}
\begin{subfigure}[b]{0.49\textwidth}
\centering
\includegraphics[scale=1]{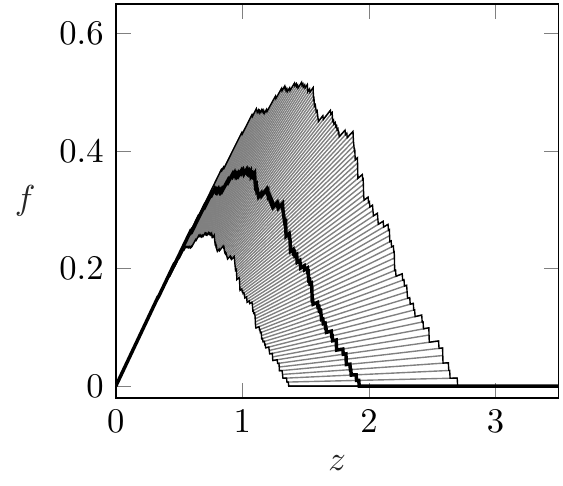}
\caption{} 
\label{fig:Tension_rho_4_lf_5.pdf}
\end{subfigure}
\caption{Mechanical response of the model in a hard device. Solid black lines, global minimum; gray lines, metastable states; (a) $\Lambda=0.8$, (b) $\Lambda\gg 1$. Parameters are $N=100$, $\lambda=1$ and thresholds follow a Weibull distribution with $\rho=4$. }\label{fig:figprob}
\end{figure}

To compute  the   equilibrium energy of the system  we need to  substitute into Eq.~\eqref{eq:ehd} the equilibrium values of the variables  $\hat y(k,z)$ and $\hat x_i(k,z)$. We obtain
\begin{equation}\label{eq:enehd}
\mathcal{H}(k,z)=\displaystyle \frac{\lambda_f}{2}\frac{ 1 - k/N }{ 1 - k/N + \Lambda} z^2+S_k.
\end{equation}
where  $S_k =  N^{-1}  \sum_{i=1}^{k} \bar{x}_i^2/2$  are  the   energies of the $k$ disrupted bonds,  and $S_0=0$.  
The tension--elongation relation for a microscopic state\footnote{In our model, due to permutation invariance,  the microscopic configuration is fully described by the number of open elements $k$.} characterized by the parameter $k$ can be now written as 
\begin{equation}\label{eq:tensionhd}
f(k,z)=\frac{\partial\mathcal{H}(k,z)}{\partial z}=\displaystyle  \frac{ \lambda_f(1 - k/N) z}{1 - k/N  +  \Lambda}.
\end{equation}
Families of equilibrium  branches parameterized by $k$ can be computed explicitly, see an example   with  $N=10$ and no disorder (homogeneous system)   in Fig.~\ref{fig:HD}.
For a  disordered system the effect of the parameter  $\Lambda$  is shown in Fig.~\ref{fig:figprob}  where  $N=100$ and the disorder is drawn from the one-parameter Weibull distribution, characterized by the probability density
\begin{equation}\label{eq:Weibull}
p(x)=\rho x^{\rho-1} exp (-x^\rho).
\end{equation}
Here $\rho$ will serve as  our second most important dimensionless parameter, inversely correlated with the 'strength' of disorder. 
 
\subsection{Response strategies}


Suppose now that the loading parameter $z$ is changing quasi-statically. As a result the system will be almost always in the state of equilibrium. However, as we have seen above, there will be many equilibrium states at each value of the loading parameter $z$, including many locally stable states. Since the equilibrium branches   of the energy $\mathcal{H}$  are defined only  for a finite range of values of the loading parameter $z$, it is still necessary   to specify  the   internal dynamics of the system which ultimately controls the choice of a particular equilibrium branch and regulates the branch switching events. To have a maximally broad picture of possible  mechanical responses we   focus below on two extreme  strategies predicting rather different  mechanical  responses \citep{PT_JMPS_2005}.

\paragraph{Marginal stability}
The  first dynamic strategy can be viewed as the vanishing viscosity limit of the overdamped viscous dynamics \citep{selinger_1991, Selinger_PRB_1991, MT_ARMA_2012}. Under this protocol,  which we will refer to as the \textit{marginal stability} (MS)   strategy/dynamics, the quasi-static loading   maintains the system in a metastable state (local minimum of the energy) till  it ceases to exist (till it becomes unstable).  Then, during an isolated switching event, the system selects a new equilibrium branch using  the steepest descent--type algorithm. This dynamic perspective can be viewed as a standard dogma  in the  macroscopic fracture mechanics of engineering structures.
 
\paragraph{Global minimum}
The second dynamic strategy  postulates  that the system is always in the \textit{global minimum} (GM) of the energy.  It implies that at each value of the loading parameter $z$, the system is able to explore the energy landscape globally and minimize the energy absolutely. Physically, this branch selection strategy  can be viewed as the zero temperature limit of the Hamiltonian equilibrium dynamics or  as a response of a  stochastic mechanical system exposed to a thermal reservoir  \citep{FRANCFORT19981319, bourdin2008variational, ET_CMT_2010}. While for macroscopic engineering structures, the global energy minimization strategy does not look too realistic, at sufficiently small scales (encountered, for instance, in cells and tissues) it may be relevant.  In such microscopic  conditions  and given that the observation is made over  sufficiently long times, thermal fluctuations can be thought as exploring sufficiently large  part of the phase space. In this sense  the GM strategy should prove  important  for the description of  low  rate  biomechanical phenomena  where the energy scale of thermal fluctuations is comparable to the size of the  energy barriers, see for instance, \cite{CT_JMPS_2017}.

\subsection{Marginal stability strategy}

To specify the MS response we first observe that each microscopic configuration exists in an extended domain of the loading parameter $z$ and that the limiting points of these domains characterize the states of marginal stability.   Define $z_k$ as the marginal stability points for the branch with $k$ broken elements. In such branch, $N-k$ elements are still holding the load, experiencing a common elongation $\bar{x}_k$ (the threshold of the $k$'th element). The force   on the bundle is therefore $(N-k)\lambda (y-\bar{x}_k)=N \lambda_f (z_k-y)$, and if we eliminate $y$ using Eq.~\eqref{eq:equiy} we obtain 
\begin{equation}\label{eq:zk_breaking}
z_k=  \frac{\lambda+1}{\lambda }\left[(1-\frac{k}{N})\frac{1}{\Lambda}+ 1 \right] \bar{x}_k.
\end{equation}
We now recall that at an elongation $x$ the expected number of broken bonds is $NP(x)$, thus only $N[1-P(x)]$ bonds carry the total load, which means that the total force is $\bar{F}(x)=N[1-P(x)]x$. The load per bond is then 
\begin{equation}\label{eq:sigave}
\bar{f}(x)=[1-P(x)]x.
\end{equation}
The average displacement $\bar{z}(x)$ at elongation $x$ can be  assessed from Eq.~\eqref{eq:zk_breaking} analytically in the limit $N\gg 1$. Then we can use the fact 
 that the empirical cumulative distribution function  $\hat{P}(x)=k/N$ is close in probability   to the cumulative distribution function   $P(x)$.  Moreover, the order statistics of $\bar{x}_k$ can be then approximated by $P^{-1} \left( k/N\right)$.  Then $\bar{x}_k \rightarrow P^{-1}(P(x))= x$ and  Eq.~\eqref{eq:zk_breaking} can be approximated by  
\begin{equation}\label{eq:zave}
\bar{z}(x)=\frac{\lambda+1}{\lambda }\left[\left[1-P(x)\right]\frac{1}{\Lambda} + 1 \right] x.
\end{equation}
If we combine  Eq.~\eqref{eq:zave} and Eq.~\eqref{eq:sigave} using  the elongation  $x$ as a parameter,  we can compute  the average force-elongation response  $\bar{f}=\bar{f}(\bar{z})$ .
\begin{figure}[ht]
\begin{subfigure}[b]{0.49\textwidth}
\centering
\includegraphics[scale=1]{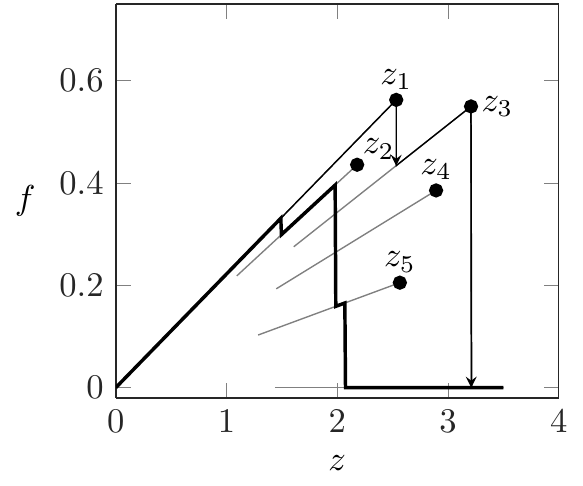}
\caption{}\label{fig:Counting_Avalanches}
\end{subfigure}
\begin{subfigure}[b]{0.49\textwidth}
\centering
\includegraphics[scale=1]{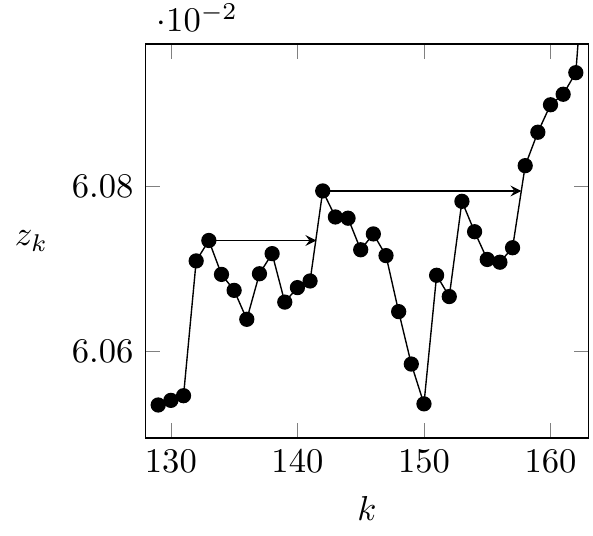}
\caption{}\label{fig:zkk}
\end{subfigure}
\caption{(a) Load path for a system with $N=5$. The thick line represent the GM response, while the thin line represent the MS response, in which the spinodal points $z_k$ are given by Eq.~\eqref{eq:zk_breaking} (b) Fluctuations of the spinodal points $z_k$   for a bundle containing $N=1000$ in the window with $129\leq k\leq 162$. The first dashed black line illustrates an avalanche of size $\Delta=9$ and the second line marks an avalanche of size $\Delta=16$. Thresholds are drawn from the Weibull distribution \eqref{eq:Weibull}  with $\rho=4$.} \label{fig:avalanches_definition}
\end{figure}


In Fig.~\ref{fig:Counting_Avalanches} we illustrate the fine structure of  a typical MS force-elongation curve using as an example a small system with $N=5$. 
We see that in this example, after the first element breaks, the system skips one metastable branch before it reaches the stable one, producing an avalanche of size two. As the load increases further, the next marginal stable endpoint $z_3$ is reached, and the system jumps to the new configuration, breaking at once two more elements and thus exhibiting an avalanche of size three. We can conclude that when the thresholds are disordered, the MS response is characterized by a series of intermittent jumps, see an example with $N=100$ in Fig. \ref{fig:Tension_rho_4_lf_5.pdf}. This figure suggests that  small   $\Lambda$ favors macroscopic fracture (collective breaking) while large $\Lambda$ produces sequential uncoordinated microfracturing as the complete synchronization is, in this case, more easily compromised by the disorder.


\subsection {Global minimization strategy}\label{sec:gmhd}

If we adopt the GM dynamics  we obtain  rather different response. 
It can be reconstructed by comparing the energies of different equilibrium configurations at a given value of $z$ and choosing the configuration that minimizes the energy. 
In contrast to the homogeneous case,  the presence of the stochastic term $S_k$ makes it challenging  to define analytically the transitions points $z_k$ separating different energy minimizing  branches.  The explicit  results can be  obtained in a simple form only at $N\gg 1$ when one  can use convenient  properties of  order statistics.

\begin{figure}[ht]
\centering
\begin{subfigure}[t]{0.49\textwidth}
\centering
\includegraphics[scale=1]{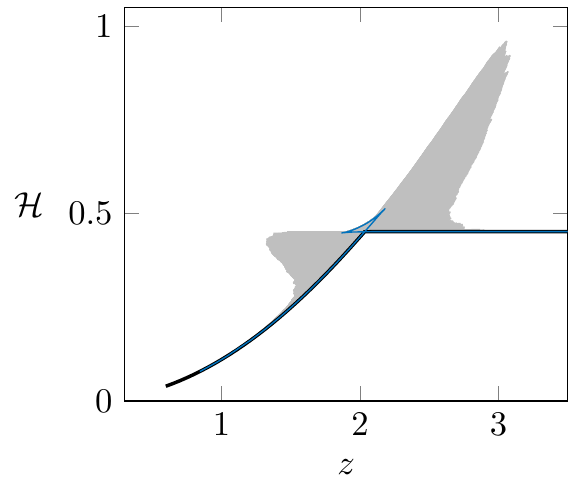}
\caption{}\label{fig:GM_Energy_Comparison_Brittle}
\end{subfigure}
\begin{subfigure}[t]{0.49\textwidth}
\centering
\includegraphics[scale=1]{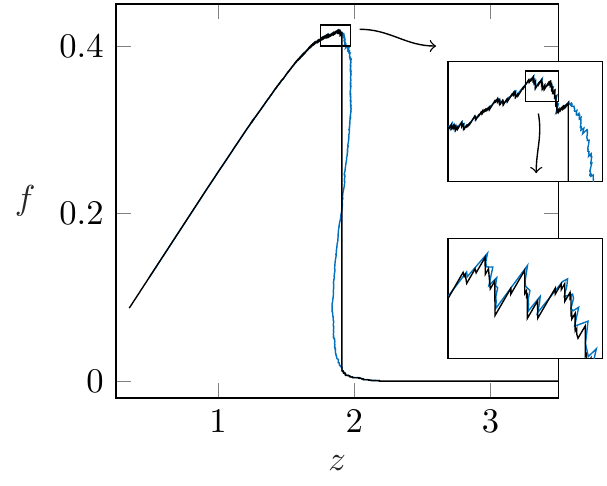}
\caption{}
\label{fig:GM_Tension_Detail}
\end{subfigure}
\caption{(a) Energy--elongation relation, (b) force-elongation. In (a) the light gray region corresponds to the metastable states; (a) and (b): the black thick line is the load path following the global minimum of the energy, the blue curve is the global minimum path as predicted by Eq.~\eqref{eq:zk_g}. Parameters: $\rho=5$, $\lambda=1$, $\lambda_f=0.4$, and $N=500$.} \label{fig:GM_Energy_Comparison}
\end{figure}

Indeed,  in the limit   $N \to \infty$ we can  approximate the sum in the definition of $S_k$ by an integral 
\begin{equation}
S_k=\frac{1}{N}\sum_{i=0}^{k} \frac{\bar{x}_i^2}{2}\approx \int_{\bar{x}_{1}}^{\bar{x}_k}  \frac{x^2}{2}dP(x),
\end{equation}
where we used that $k/N\rightarrow P(\bar{x}_k)$, and    identified the increment   $dP(x)$ with $1/N$.
The ensuing asymptotic expression   for the  energy, Eq.~\eqref{eq:enehd} can be written in the form 
\begin{equation}
\mathcal{H}(x,z)=\frac{\lambda_f (1-P(x))}{1-P(x)+\Lambda}\frac{z^2}{2}+\int_{0}^{x} p(x')\frac{x'^2}{2}dx'.
\end{equation}
From the equilibrium equation $\partial \mathcal{H}(x,z)/\partial x =0$  we obtain the desired relation $\bar z(x)$.
\begin{equation}\label{eq:z_g-ave}
\bar z=\left(1-P(x)+\Lambda \right)\frac{x}{\sqrt{\lambda_f \Lambda}}.
\end{equation}
%
To see how relevant  such approximation  may be at finite $N$ we rewrite the asymptotic  relation  \eqref{eq:z_g-ave}, in a discrete form, by assuming that the continuous variable $x$ takes only discrete values  $\bar{x}_k$ 
\begin{equation}\label{eq:zk_g}
z_k=\left(1-\frac{k}{N}+\Lambda \right)\frac{\bar{x}_k}{\sqrt{\lambda_f \Lambda}}.
\end{equation}
We can then substitute  \eqref{eq:zk_g} into  \eqref{eq:enehd}  and reconstruct  the GM response.  In Fig.~\ref{fig:GM_Tension_Detail}, we compare   for the case $N=500$ the  results of direct  numerical  minimization of the energy with the theoretical  prediction based on \eqref{eq:zk_g}. The  overall agreement is very good, even at finer scales, even though,  the approximate theory does not capture every single fluctuation. 

\begin{figure}[ht]
\centering
\includegraphics[scale=.9]{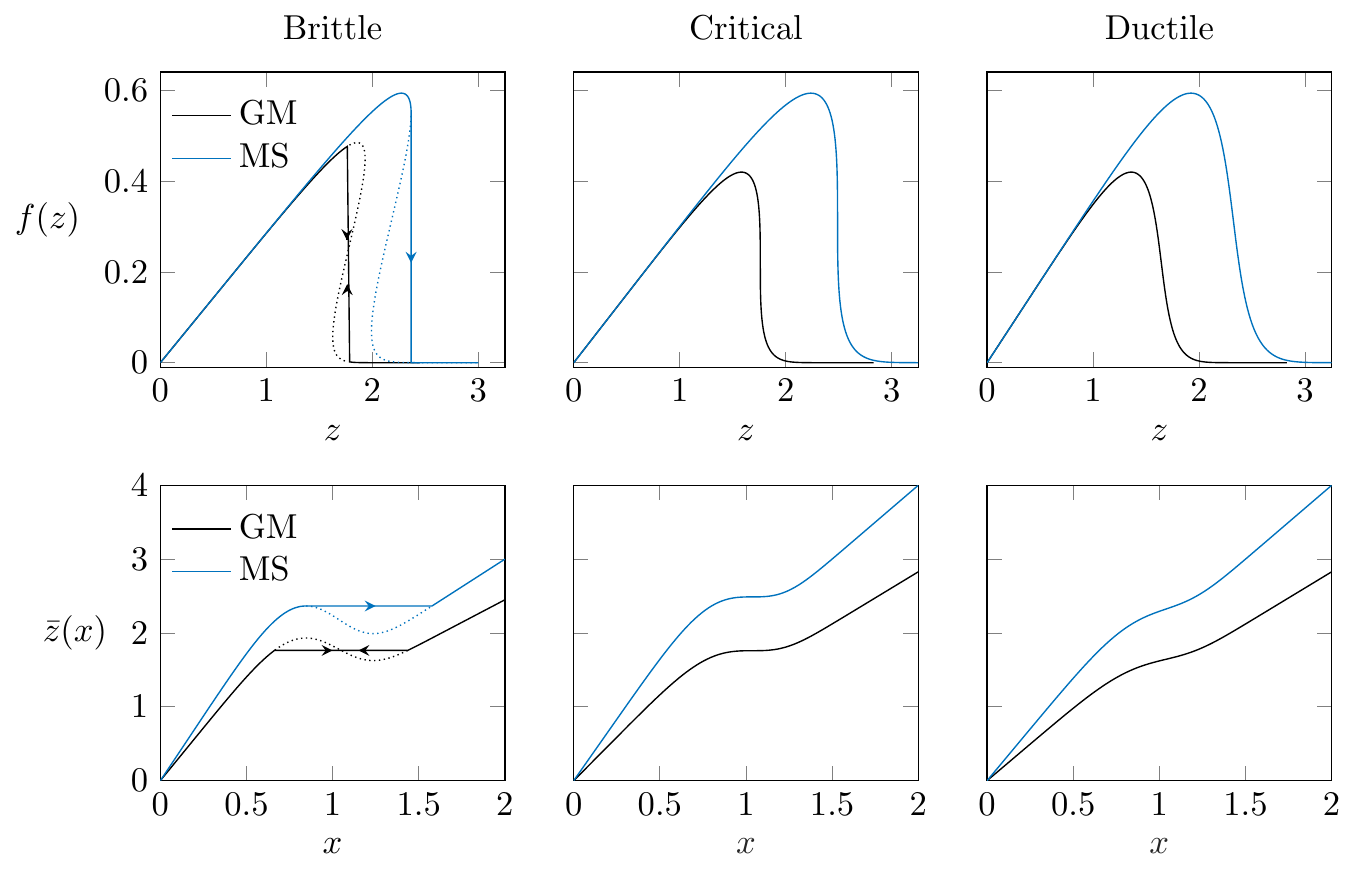}
\caption{First row: force elongation, second row: external elongation as a function of internal variable $x$. The blue curves correspond to the unbinding path under  the MS dynamics; The black curves correspond to the GM dynamics.} \label{fig:Average_properties}
\end{figure}

The  $N \to \infty$ limits for  MS and GM paths are compared  in Fig.~\ref{fig:Average_properties}. It illustrates  the important role played by the  parameter  $\Lambda$.   At a fixed disorder, the system fractures  gradually when $\Lambda$ is large, and we identify this regime with  \emph{ductile} fracture. Instead, at small values of $\Lambda$ we observe a finite abrupt discontinuity and associate such regimes with  \emph{brittle} fracture. 
Since our averaged force-elongation curves are obtained from the parametric equations for $f(x)$ and $z(x)$, whenever  we have a discontinuity in $f(z)$   we also have a discontinuity in $z(x)$.  In other words, if  at a fixed elongation $z$, there are multiple configurations   satisfying \eqref{eq:z_g-ave} or \eqref{eq:zave},  we deal with the  brittle regime.

\subsection{Brittle to ductile transition}

To distinguish between brittle and ductile responses in the framework  of   MS dynamics,
we   need to check if there are  roots of the equation  $ \partial \bar{z} (x)/ \partial x=0$, where $\bar{z}(x)$ is given by Eq.~\eqref{eq:zave}, corresponding to a local maximum.  If  there is such a  root $x=x_{*}$ it solves the equation
\begin{equation}\label{eq:mxz}
1-P(x)-p(x)x+\Lambda =0.
\end{equation}
The existence of solutions of  Eq.~\eqref{eq:mxz} is then  a fingerprint of   collective debonding events, see   Fig.~\ref{fig:Average_properties}. 
The critical regime, separating brittle from ductile responses,   corresponds to the parameter choice  when the solution of  Eq.~\eqref{eq:mxz} is an inflection point. Such $x=x_c$  should also solve  $\partial^2 \bar{z} (x)/\partial x^2=0$, which can be rewritten as 
\begin{equation}\label{eq:inflection_marginal}
-2 p(x) -p'(x) x=0.
\end{equation}
To summarize,   the critical manifold  in the space of parameters   can be found by eliminating $x$ from Eq.~\eqref{eq:mxz} and Eq.~\eqref{eq:inflection_marginal}.

Note next that   whenever the system is brittle under  the MS  dynamics,  it is also brittle under  the GM dynamics.  However, in the latter case the system does not reach the maximum on the curve $\bar{z} (x)$ (spinodal point),    because the  energy minimizing configuration is switched earlier and the point of  collective debonding should be chosen instead by the Maxwell construction, see  Fig.~\ref{fig:Average_properties}.  
According to  Eq.~\eqref{eq:z_g-ave},  the condition for the   local maximum of $\bar{z} (x)$ is again given by \eqref{eq:mxz}.
 Similarly, the condition for an inflection point is again \eqref{eq:inflection_marginal}.
 Hence, the   critical  manifold,  separating brittle from ductile regimes in the space of parameters, is  independent of the loading protocol.
%
%
%
%
%
%

The location of the boundary separating brittle and ductile regimes
depends on the strength of the disorder (the parameter $\rho$ for the Weibull distribution)
and on the dimensionless parameter 
$\Lambda$
which we interpret as a measure of the structural   rigidity of the system \citep{Merkel6560, Vermeulen_PRE_2017, crapo1979structural, Kim_NatPhys_2019}.  When $\Lambda$ is large, meaning that either $\lambda_f$ is large or $\lambda$ is small, individual breakable elements interact weakly. 
In fact,  the limit $\Lambda \to \infty$ can be associated with the  jamming  threshold (isostatic point)  beyond which the rigidity is completely lost \citep{Goodrich2014}. Instead, when $\Lambda \to 0$ the system can be viewed as over-constrained \cite{Vitelli_PNAS_2016}. In some sense, the dimensionless parameter $\Lambda$ is similar to the effective Poisson's ratio $\nu$, which is by itself positively correlated with the ratio $\kappa/\mu$ where $\kappa$ is the effective bulk modulus and $\mu$ is the effective shear modulus. Both these dimensionless parameters decrease with increased connectivity of the lattice/network,  which diminishes fracture toughness and favors brittleness. Instead they increase with denser packing, which enhances fracture toughness and facilitates ductility \cite{Greaves2011, richard2021brittle, Deng_JAP_2018, ma12152439}

\begin{figure}[ht]
 \centering
\includegraphics[scale=1]{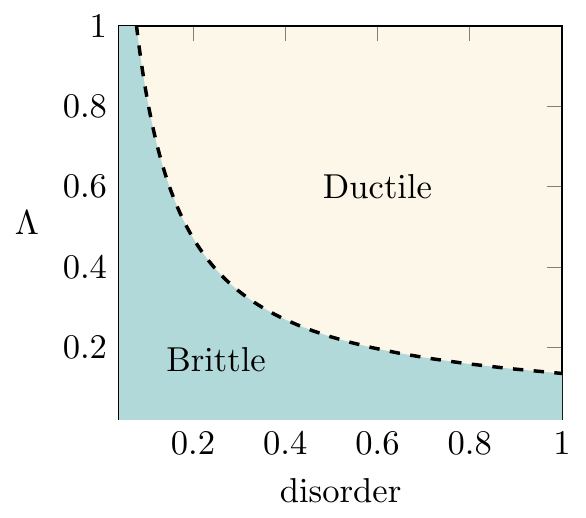}
\caption{The phase diagram constructed based on Eq~\eqref{eq:lfw}.  We used the variance of the Weibull distribution, $\text{Var} (\rho) = \Gamma\left(1+ 2/\rho\right)-\left[\Gamma\left( 1+ 1/\rho\right) \right]^2$, as the measure of the 'strength' of  disorder. The 'malleability' parameter  $\Lambda$  is  viewed as inversely correlated to  the system's rigidity. A closely related but differently parameterized empirical phase diagram  has been recently obtained in \citep{david2021finitesize}.} \label{fig:AthermalPD}
\end{figure}

To illustrate these results, we use again our one-parameter Weibull distribution of thresholds, characterized by a single shape parameter $\rho$  which  can be viewed as inversely proportional to the variance of the disorder. One can show that in this case the boundary between the brittle and ductile regimes is given by the explicit relation  
\begin{equation}\label{eq:lfw}
\Lambda = \rho \exp{ (- 1/\rho -1)}.
\end{equation}
According to \eqref{eq:lfw} ductility can be achieved by decreasing the degree of interaction/connectivity among individual binders measured by any  rigidity measure negatively correlated, say, inversely proportional \cite{BRT_PRL_2020},  to our parameter $\Lambda$.  Indeed,    the bigger the $\Lambda$, the less sensitive the binders are to a change in the configuration of their neighbors.  Since disorder tends to desynchronize the bonds,  Eq.~\ref{eq:lfw}  predicts a transition from brittle response (correlated microfailure) at small disorder to ductile response (uncorrelated microfailure) at large disorder. The corresponding phase diagram in the space  rigidity--disorder is presented in Fig.~\ref{fig:AthermalPD}.

%

\section{Empirical avalanche distribution}

So far  we were mostly concerned with the averaged behavior,  neglecting fluctuations. Now, we change focus and turn to the statistics of avalanches as the system is quasi-statically loaded in the hard loading device. 

%
%
\subsection{Marginal stability strategy}

The occurrence of avalanches under MS dynamics can be predicted if we know the sequence of external elongations $\{z_k\}$, defined as the instability points for the branches with $k-1$ broken elements and given by  \eqref{eq:zk_breaking}.
For an avalanche  of size $\Delta$ to start when the $k$th fiber is about to break,  two conditions must be satisfied. The first condition is called the \textit{forward condition} and it states that $\Delta-1$ fibers must fail after the breaking of the $k$th fiber. The second condition, called the \textit{backwards condition}, ensures that the avalanche starts indeed with the breaking of the $k$th fiber and it is not a part of a bigger one.  These conditions will be used in our numerical experiments in the analytical form
\begin{enumerate}
\item $z_{k+j}\leq z_{k},\,{\textstyle \mbox{{for}}\:}\, j=1,2,\dots,\Delta-1$ and $z_{k+\Delta}>z_{k}.$ (forward condition)
\item $z_{j}\leq z_{k}$, for all $j<k$, (backwards condition)
\end{enumerate}


\paragraph{Numerical experiments.} To reveal the  response  of the system we conducted a series of numerical experiments varying the parameters of the system and the realizations of disorder. Our experiments  unveiled  the existence of the two  main fracture regimes (brittle and ductile). They  also showed that the range of scaling blowd up at the boundary between the two regimes suggesting criticality.

\begin{figure}[ht]
\includegraphics[width=\textwidth]{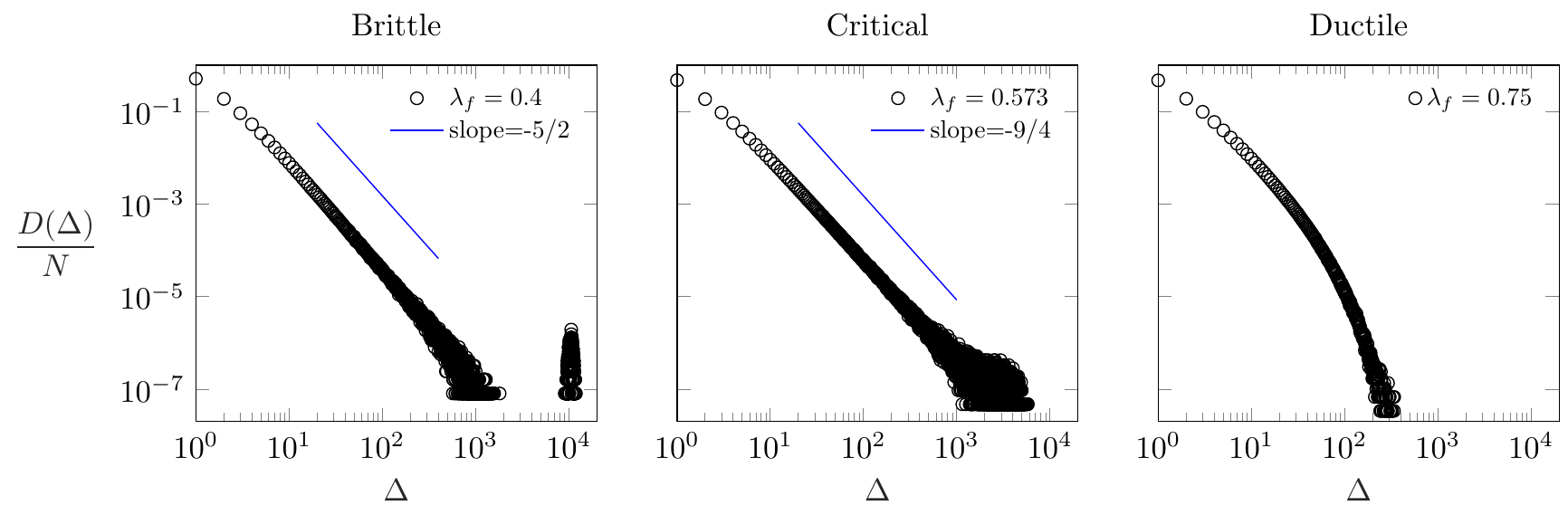}
\caption{Avalanche distribution $p_a(\Delta) =D(\Delta)/N$ where $D(\Delta)$ is the number of avalanches with the  size $ \Delta$.  We used  MS protocol for the system with $N=20 000$, $\lambda=1$ and thresholds following a Weibull distribution with $l=1$ and $\rho=4$. The distribution is averaged over 10000 realizations of the disorder. }\label{fig:av}
\end{figure}

More specifically we were   interested in the expected number of bursts of a particular size $\Delta$.  In Fig.~\ref{fig:av}, we present the   numerically computed  avalanche distribution $p_a(\Delta)$  in the brittle ($\lambda_f=0.4$), critical ($\lambda_f=0.573$) and ductile ($\lambda_f=0.75$) regimes under the   MS dynamics.   In the brittle regime, we found the power law type distribution with
$$p_a(\Delta) \sim \Delta^{-\alpha}.$$ 
The exponent $\alpha=5/2$ turned out to be  the same one as the one  found in the classical force-controlled FBM
\citep{hansen2015fiber, Hansen:262741, hansen, Kloster, Pradhan:2010aa}.  However, the obtained distribution is only super-critical because of the presence of the peak, representing system-size events.

In the critical regime, we also observed the power law distribution but now with the exponent $\alpha=9/4$, which was not known in the classical FBM. Finally, in the ductile regime, the distribution is of power law type only for very small events but then exhibits an exponential decay suggesting  Gaussian distribution.


\subsection{Global minimum strategy}

Under the  GM  dynamics  the response  reduces to the global  minimization of the energy \eqref{eq:enehd}. Suppose that an avalanche event happens at an elongation $z=\zeta$, in a state where the maximally stable configuration has $k$ bonds that are broken. If $\Delta$ further bonds are broken in the avalanche, the next stable configuration must satisfy
$\mathcal{H}(\zeta , k) = \mathcal{H}(\zeta , k+\Delta).$
Moreover, we have to ensure that the configurations $k$ and $k+\Delta$ are indeed the ones that minimize the energy at $\zeta$. Thus the energy of the configurations between the two states $k, \, k+\Delta$ have to be greater than $\mathcal{H}(\zeta , k)$, which means
$\mathcal{H}(\zeta , k) < \mathcal{H}(\zeta , k+j), \text{ with } j=1,\dots,\Delta-1.$
Let $E_i (\zeta)\equiv \mathcal{H}(z = \zeta, i)$ be the sequence of the energies of the the system at fixed elongation $\zeta$ corresponding to different configurations $i$. The conditions for the avalanche of size $\Delta$ starting at $k$ can be then formulated in terms of $E_i (\zeta)$ as,

\begin{enumerate}
\item $E_{k} (\zeta) < E_{k+i} (\zeta)$, for $i=1,\dots,\Delta-1$ and $E_{k} (\zeta)\geq E_{k+\Delta}(\zeta)$ (forward condition)
\item $E_{k} (\zeta) < E_{j} (\zeta)$, for $j<k$ (backwards condition)
\end{enumerate}

\begin{figure}[ht]
\centering
\includegraphics[width=\textwidth]{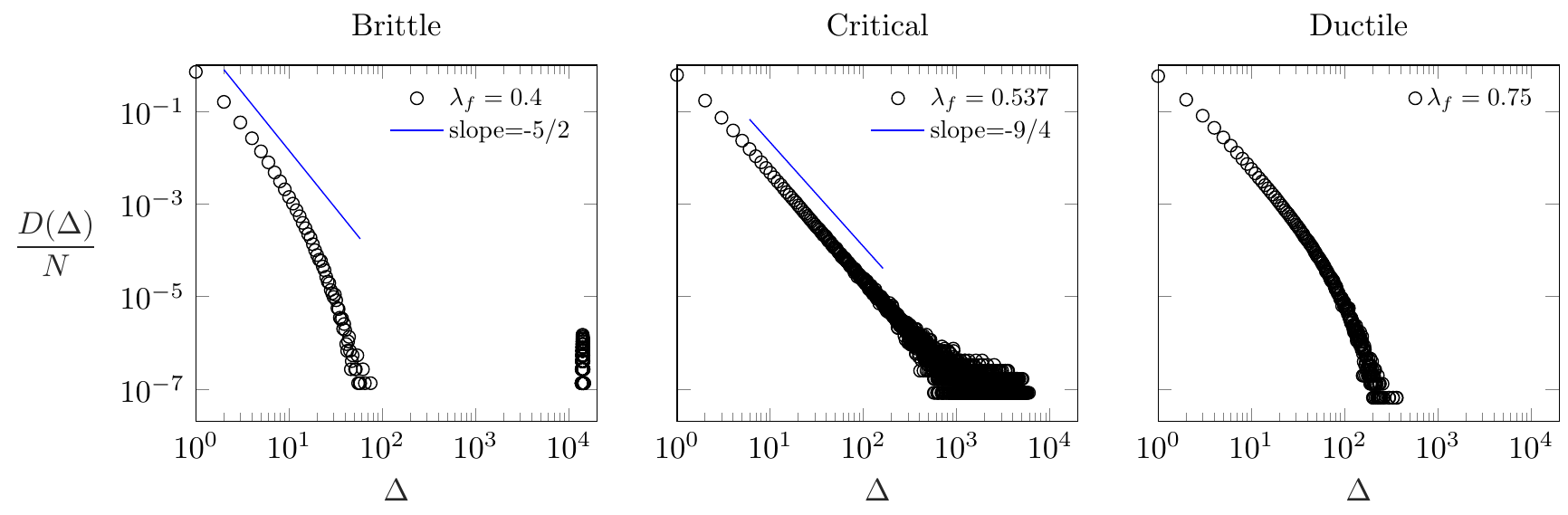}
\caption{Avalanche distribution $p_a(\Delta) =D(\Delta)/N$ where $D(\Delta)$ is the number of avalanches with the  size $ \Delta$.  We used  the GM protocol  for the system with  $N=20000$,  $\lambda=1$ and thresholds following a Weibull distribution with $l=1$ and $\rho=4$. The distribution is averaged over 10000 realizations of the disorder.}
\label{fig:GM_Avalanches}
\end{figure}

\paragraph{Numerical experiments.} Using these conditions we can  obtain numerically the force-elongation curve and study the statistics of avalanches. Our numerical experiments,  summarized  in Fig.~\ref{fig:GM_Avalanches}, show that the overall structure for the breaking sequence along the GM path is similar to the one we obtained in the case of the MS dynamics.
 When the $k$'th bond is about to break to ensure the global minimization of the energy, the elongation per fiber is given by Eq.~\eqref{eq:zk_g}. Note   that Eq.~\eqref{eq:zk_g} was obtained for $N\gg 1$. This means that for large clusters we can compute the avalanche distribution in a similar way as in the out of equilibrium case. Moreover, one should then expect that the statistical features of equilibrium (GM) and nonequilibrium (MS) avalanche distributions would be the same. Our numerical experiments show that, indeed,  the equilibrium avalanche statistics in the critical and ductile regimes are identical to the ones in the out-of-equilibrium system. However, as we can see in Fig.~\ref{fig:GM_Avalanches}, the statistics for the brittle regime are not the same. While in the case of MS dynamics we see a power law with a peak, in the global minimization response we see a power law with exponential decay followed by a peak. The reason is that in the latter case the system collectively fractures before it reaches the state where the stiffness diverges.  

Our numerical experiments  suggested the direction of the development of the analytical  theory which is presented in the next sections.  Such theory  must  justify the existence of the two  different fracture regimes (brittle and ductile) and it is expected to show that equilibrium and non-equilibrium exponents agree. It should also be able to show that the range of scaling blows up at the boundary between the two regimes suggesting full scale criticality  on the BDT  boundary.  

\section{Avalanche as a random walk}

To justify analytically the empirical avalanche distributions shown in Fig.~\ref{fig:av} and Fig.~\ref{fig:GM_Avalanches} we first show that the process of propagating fracture can be represented by a (biased) random walk, see for the initial insights \cite{Sornette_1992, Pradhan:2010aa, sornette2006critical}. Each step of the random walk represents the breaking of a single element, and therefore such random walk has a constant step size. Because of the random nature of the thresholds $\bar{x}_k$ an avalanche event can be seen as a return (first passage) time problem for a  random walk.
To compute the probability of an avalanche of size $\Delta$ happening at configuration $k$ is then equivalent to computing the probability of the random walk $z_k$ return after $\Delta$ steps.



\paragraph{MS dynamics.} 

To compute the probability of an avalanche of size $\Delta$ along the  MS path which starts with the breaking of the $k$th element   we need to know the probability distribution of the \emph{small}  length increments $\delta z (x)= z_{k+1}-z_{k}$, where we assumed that at the point of  instability  $z_k=z$  the configuration is $\bar{x}_k = x$.    In the limit  $N \gg 1$ it will be more convenient to deal with the  \emph{finite} length increments (jumps)  $[[Z]] =N \delta z  (\lambda/(\lambda+1)) \Lambda$. We can use \eqref{eq:zk_breaking} to write 
\begin{equation}
[[Z]]= (1-k/N+\Lambda)N\delta x
- x,
\label{delta}
\end{equation}
where    $\delta x=\bar{x}_{k+1} - x$ are the \emph{ small }increments of the threshold values. The probability distribution  of $\bar{x}_{k+1}$, under the condition that $\bar{x}_k = x$,  can be written as 
\begin{align}
\label{delta1}
\rho(\bar{x}_{k+1})   &= (N-k)\left[ \frac{1-P(\bar{x}_{k+1})}{1-P(x)}\right]^{N-k-1} \frac{p(\bar{x}_{k+1})}{1-P(x)}.
\end{align}
 Using \eqref{delta} and the fact that  $N \gg 1$  we can now rewrite \eqref{delta1}  as  the probability distribution $\rho([[Z]])$.  Indeed, in the limit   $N \to \infty$, we have  $k/N \approx P(x)$,   $p(\bar{x}_{k+1})\approx p(x)$.   Moreover,  since  $\bar{x}_k \to P^{-1}(k/N)$, we can use the approximation  $\bar{x}_{k+1}-\bar{x}_{k} \to P^{-1}((k+1)/N)-P^{-1}(k/N) \sim 1/N$. Then, given that  $\delta x \ll x$, we can write  
\begin{align}\label{eq:PoissonLimit}
\left[ \frac{1-P(x + \delta x)}{1-P(x)}\right]^{N-k-1} & \approx
\exp\left[ -p(x) \left( \frac{[[Z]]+ x}{1-P(x)+\Lambda}\right)\right].
\end{align}
Hence 
\begin{equation}\label{eq:pdfdeltaZ}
\rho ([[Z]])
  = \left \{ \begin{array}{l l}
   \frac{p(x)}{1-P(x)+\Lambda}   \exp\left[ - \frac{p(x)}{1-P(x)+\Lambda} \left([[Z]] +x\right)\right] , & [[Z]]+x \geq0, \\
   0, & [[Z]]+x <0. 
 \end{array}
 \right.
\end{equation}
This  distribution defines the asymmetric random walk with  the average step,
\begin{equation} \langle [[Z]]\rangle= (1-P(x)+\Lambda-x \, p(x))/(p(x)),
\end{equation}
which is non-zero, except at the critical or spinodal points, see \eqref{eq:mxz}.
To characterize the random walk fully, we would also need to know the variance 
\begin{equation} \sigma_{[[Z]]}^2 =  (1+\Lambda-P(x))^2/ p(x)^2.
\end{equation}
Then the step bias of the ensuing random walk is 
\begin{equation}\label{eq:bias_ms}
b \equiv \frac{\langle [[Z]]\rangle}{\sigma_{[[Z]]} } = 1-\frac{x \,p(x)}{1+\Lambda-P(x)}=1-g(x),
\end{equation}
where 
\begin{equation}
g(x)=\frac{ p(x)\, x}{1-P(x)+\Lambda}.
\end{equation} 
Note that $b=0$ in spinodal and critical regimes, where $g(x)=1$ (see Eq.~\eqref{eq:mxz}), and $b>0$  otherwise.

\paragraph{GM dynamics.} 
%
%

Similarly, in the context of GM dynamics we can consider the sequence $E_k$ as a   biased random walk with unitary step sizes.
As before, we will focus on the  computation of the asymptotic probability distribution for   the \emph{finite} energy increments $[[E]](x) = N(E_{k+1}(x)-E_k(x)) $, where  $E_k$ corresponds to  $\bar{x}_k = x$.  Given that   $N \gg 1$ we can approximate $x$ using  Eq.~\eqref{eq:zk_g} and  then use  Eq.~\eqref{eq:enehd} to write,
\begin{equation}\label{eq:DeltaE}
\begin{split}
[[E]]  
& = -\frac{1}{1- (1-k/N+\Lambda)^{-1} N^{-1}} \frac{x^2}{2}+  \frac{\bar{x}_{k+1}^2}{2}
 \approx - \left[ 1+ \frac{1}{(1-k/N+\Lambda)N}\right]\frac{x^2}{2} +  \frac{\bar{x}_{k+1}^2}{2}.
\end{split}
\end{equation}
%
Next, we introduce  a new random variable $\xi = x^2/2$ with cumulative distribution $F(\xi)$ and  probability density $f(\xi)$.
Given that $\xi_k = \xi$, the distribution function for $\xi_{k+1}$ can be written as
\begin{align}
 \rho(\xi_{k+1}) = (N-k)\left[ \frac{1-F(\xi_{k+1})}{1-F(\xi)}\right]^{N-k-1} \frac{f(\xi_{k+1})}{1-F(\xi)}.
\end{align}
We can then use \eqref{eq:DeltaE} to  obtain the probability   density $\rho([[E]])$.  In the limit $N \to \infty $ we can write
 \begin{equation}
\rho ([[E]]) = (N-k) \left[ \frac{1-F(\xi +\delta\xi)}{1-F(\xi)}\right]^{N-k-1} \frac{f(\xi_{k+1})}{1-F(\xi)}
\end{equation}
 and, since the increments  $\delta\xi = N^{-1}[[E]] + [(1-k/N+\Lambda) N]^{-1}\xi \sim 1/N$ are \emph{small} $\delta \xi  \ll \xi $, we  can use the approximation 
\begin{align}
\left[ \frac{1-F( \xi +\delta\xi)}{1-F(\xi)}\right]^{N-k-1}
 \approx \exp\left[ -f(\xi) \left([[E]] +\frac{\xi}{1-F(\xi)+\Lambda}\right)\right].
\end{align}
Finally, using the fact that  $f(\xi_{k+1})\approx f(\xi)$, we obtain the desired distribution
\begin{equation}
\rho ([[E]])
  = \left \{ \begin{array}{l l}
   N\, f(\xi)   \exp\left[ - f(\xi) \left([[E]] +\frac{\xi}{1-F(\xi)+\Lambda}\right)\right] , & [[E]]+ \frac{\xi}{1-F(\xi)+\Lambda} \geq 0, \\
   0, & [[E]] + \frac{\xi}{1-F(\xi)+\Lambda}<0. 
 \end{array}
 \right.
 \label{energy}
\end{equation}
Since $\bar{x}_i\geq 0$, the   probability distributions for $\xi = x^2/2$ can be  expressed through the known distribution for  $x$, in particular,  $F(\xi) = P(x)$, and $f(\xi) = \frac{2}{ x}p(x)$. Therefore we can rewrite  \eqref{energy} as
\begin{equation}
\rho ([[E]])
  = \left \{ \begin{array}{l l}
   N\, \frac{2 p(x)}{x}   \exp\left[ - \frac{2p(x)}{x} \left([[E]] +\frac{x^2/2}{1-P(x)+\Lambda}\right)\right] , & [[E]] + \frac{x^2/2}{1-P(x)+\Lambda}\geq 0, \\
   0, & [[E]]+ \frac{x^2/2}{1-P(x)+\Lambda} <0. 
 \end{array}
 \right.
\end{equation}
For the ensuing random walk we  can now compute the mean 
\begin{equation}\langle [[E]]\rangle=  x(1-g(x))/(2N p(x)),
\end{equation}
 and the variance
\begin{equation}\sigma_{[[E]]}^2 =  x^2/(2N p(x))^2.\end{equation}
The corresponding step bias is then
\begin{equation}\label{eq:bias_gm}
b\equiv\frac{\langle [[E]]\rangle}{\sigma_{[[E]]} } = 1-g(x),
\end{equation}
which is exactly the same expression as in the case of MS dynamics.


Given that  we  mapped  avalanches on one dimensional random walks with constant step size, the statistics of avalanches can be obtained from the solution of the general \emph{first passage} time problem for such stochastic processes~\cite{metzler2014first}. Our return problem is then equivalent of the first passage problem with  the "return point"   chosen to coincide with the origin.  

Indeed, consider an abstract 1D random walk of this type  and denote the probability of taking a step in the positive direction  by  $p$ and in the negative direction by $q =1-p$. For such random walk, the average advance is
\begin{equation}
\langle [[r]] \rangle  = p-q,
\end{equation}
 the standard deviation is 
\begin{equation}
\sigma_{[[r]]} =\sqrt{4pq }. 
\end{equation}
 Then for  the step bias relative to the standard deviation,   we obtain
\begin{equation}\label{eq:StepBias}
\frac{\langle [[r]] \rangle}{\sigma_{[[r]]}} =\frac{p-q}{\sqrt{4pq }}.
\end{equation}
For a general  random walk,  satisfying the forward condition is equivalent to  solving  a special case of the  gambler's ruin problem in the interval $(0,\infty)$ with an absorbing barrier at 0.
The amount of steps before the return is exactly  the avalanche size $\Delta$.

To find the avalanche distribution,  we define the probability $\psi_{1,n}$  that our general  random walk  initiated  at $r=1$   returns   at $  r=n$. The solution of this  classical problem is known \cite{feller_Probability}  
\begin{equation}
\psi_{1,n}=\frac{1}{n} \binom{n}{(n+1)/2} p^{(n-1)/2}q^{(n+1)/2}.
\end{equation}
We now assume that $n \gg1 $  and associate with $n$  the  avalanche of size $\Delta$.   Then the avalanche distribution is  $p \, \psi_{1,\Delta}$, where the $p$ factor  ensures that the first step is in the positive direction. We obtain
\begin{equation}
p \,\psi_{1,\Delta}\approx \sqrt{1/2\pi} \, \Delta^{-3/2} \, (4pq)^{\Delta/2}\sqrt{p/q}.
\end{equation}

We can now associate our general random walk with the discrete failure process under either  MS or  GM protocols. We recall that the relative bias is the same under both conditions, and hence we can use the identification
\begin{equation}
\frac{p-q}{\sqrt{4pq }} =  \frac{\langle [[E]]\rangle}{\sigma_{[[E]]} } =\frac{\langle [[Z]]\rangle}{\sigma_{[[Z]]} } = 1-g(x).
\end{equation}
Since we are mainly interested in the case of  \emph{small} bias $b$, as the main avalanche statistics is acquired near either spinodal or critical points,  we can write $p=(1+b)/2$ and $q=(1-b)/2$. Therefore  
\begin{equation}
 \frac{p-q}{\sqrt{4pq }}=
 \frac{b}{1-b^2} \approx b.
\end{equation}
We have thus  used the  forward condition.  The backwards condition states  that  the zero level  is not  reached before the end of the avalanche. Under the assumption that $n \gg1 $  this means that  the time-reversed process never returns. The probability of no return for such random walk is \cite{feller_Probability, hansen2015fiber}
\begin{equation}
p_{bc} = \vert p-q\vert = b.
\end{equation}
Bringing these results together, we can present  the probability of having an avalanche of size $\Delta$ in the form 
\begin{equation}
\Psi(\Delta)  =p \,\psi_{1,\Delta} \, p_{bc}=  \frac{1}{\sqrt{2\pi}} \, \Delta^{-3/2} \, b \,  e ^{\log(1-b^2)\Delta/2} \sqrt{\frac{1+b}{1-b}} 
\approx  \frac{1}{\sqrt{2\pi}} \, \Delta^{-3/2} \, b \,  e ^{-b^2\Delta/2},
\end{equation}
where we used the small bias approximations  $\log (1-b^2) \approx -b^2$, and 
$\sqrt{(1+b)/(1-b)} \approx 1$.

In both MS and GM dynamics, we deal with cases where the parameter $b$ depends on the value of the elongation $x$ at which breaking begins. To  account for this parametric dependence we use the notation  $ \Psi(\Delta;x)$. Since the number of elements with thresholds in  $(x,x+dx)$ is $N p(x)dx$,  the number of avalanches of size $\Delta$ starting inside $(x,x+dx)$ is $ \Psi(\Delta;x) N p(x)dx $. Then the distribution of avalanches in the range  $(x_{i}, x_{s})$, close to either spinodal or critical point is
\begin{equation}\label{eq:distr}
p_a(\Delta) \equiv \frac{D\left(\Delta\right)}{N}= {\displaystyle \frac{1}{\sqrt{2\pi}} \, \Delta^{-3/2}\int_{x_{i}}^{x_{s}}} \phi(x) e^{-h(x)\Delta }dx,
\end{equation}
where  $\displaystyle {\phi(x)=\left[1-g(x)\right]p(x)}$ and $ h(x)=[1-g(x)]^2$.  The distribution \eqref{eq:distr} has the same structure as the one obtained in the classical fiber bundle model loaded in a soft device \citep{hansen2015fiber,hansen, Kun_EPL_2010}. The new element is the appearance of the   parameter  $\Lambda =\lambda_f(\lambda+1)/\lambda$, which   can differentiate between brittle and ductile behaviors. Numerical simulations, Fig.~\ref{fig:av} and Fig.~\ref{fig:GM_Avalanches}, suggest different exponents in different ranges of $\Lambda$   and our asymptotic  analysis of Eq.~\ref{eq:distr} in the next section  will reveal  the origin of these differences. 
 
\section{Avalanche distribution:  asymptotic results}

Since both MS and GM random walks have the same bias $b(x)$, the corresponding asymptotics for the avalanche distribution can be expected to be the same in the critical and the ductile regimes. In the brittle regime, the MS and GM dynamics can be expected to produce different types of system spanning events, spinodal and equilibrium (Maxwellian), respectively, and therefore these two cases would have to be treated separately.

Using the fact that $\Delta $ is large, we can approximate the integral in Eq.~\eqref{eq:distr} using  Laplace's method around  the global minimum of the function $h(x)$. If the minimum at, say $x=x_0$, exists, then for large $\Delta$ the main contribution to the integral will come from the vicinity of $x_0$.  To find this point we need to solve the equation
$
h'(x)=2g'(x)(1-g(x))=0.
$
There are two classes of solutions of this equation corresponding to either $g(x_0)=1$ or $g'(x_0)=0$. This leads to three possibilities,
\begin{enumerate}
\item $g(x_0)\neq 1$ and $g'(x_0)= 0$,
\item $g(x_0)=1$ and $g'(x_0)= 0$,
\item $g(x_0)=1$ and $g'(x_0)\neq 0$.
\end{enumerate}
These three types of solutions  produce our three regimes:  ductile, critical and brittle, respectively.  

\begin{figure}[ht]
\centering
\includegraphics[width=\textwidth]{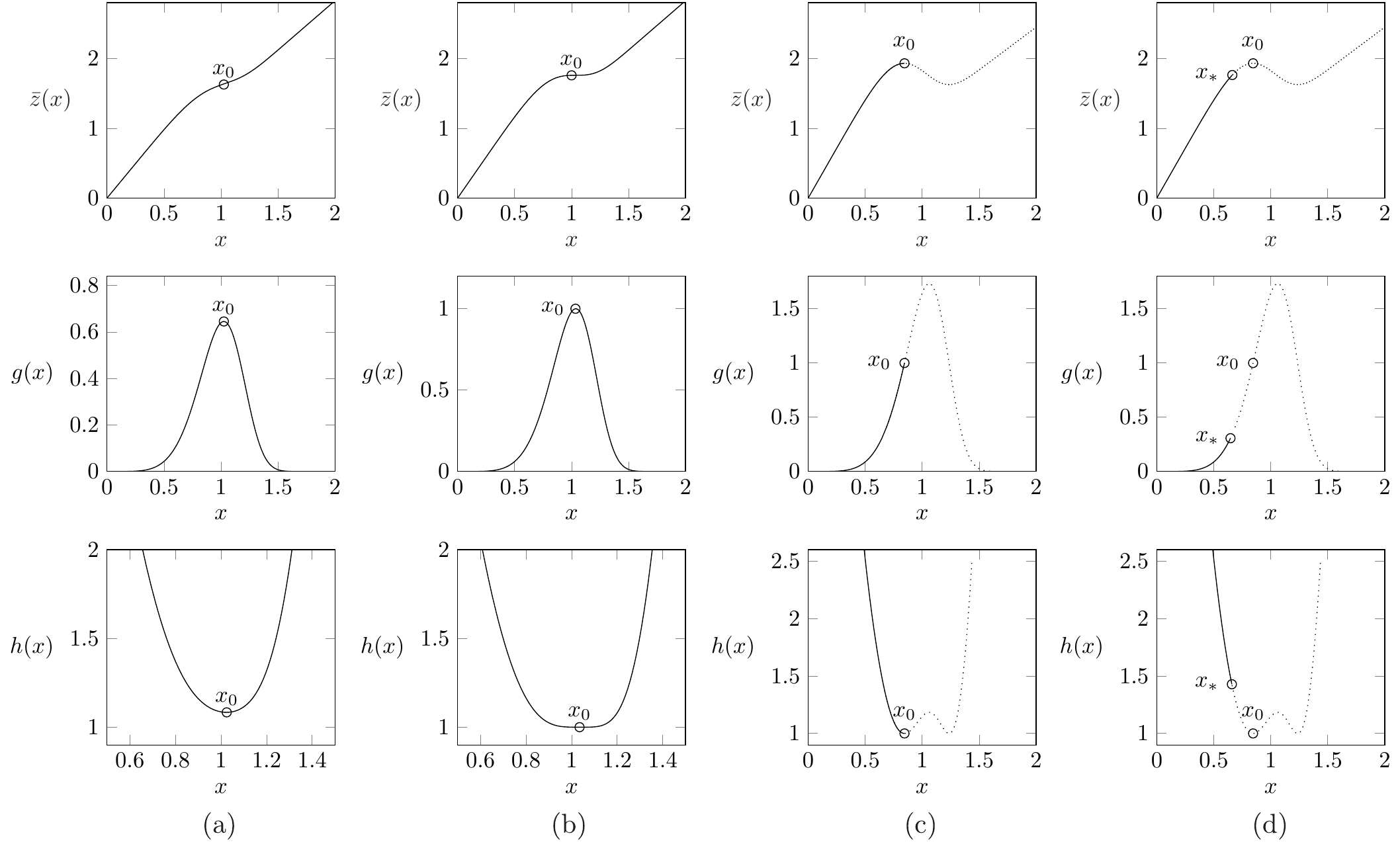}
\caption{Behavior of the functions $z(x)$, $g(x)$ and $h(x)$ when the system is in: (a) the ductile regime; (b)  the critical regime; (c)   the brittle regime under  the MS dynamics; (d)  the brittle regime under the GM dynamics.} \label{fig:Asymptotics_Ductile}
\end{figure}

Indeed, the condition $g(x_0)=1$ is equivalent to,
\begin{equation}
\label{brittle}
p(x_0)x_0=1-P(x_0)+\Lambda,
\end{equation}
which is our condition of brittleness  based on the averaged  description, see  Eq.~\eqref{eq:mxz}. The condition $g'(x_0)=0$ can be rewritten as
$
 2/x_0+ p'(x_0)/p(x_0)=0,
$
and   is therefore equivalent to the condition $\partial^2 \bar{z}(x)/\partial x^2=0$ indicating   an inflection point on the $z(x)$ curve. In the absence of the minimum of $h(x)$, we have $g(x_0)\neq 1$, and then the condition $g'(x_0)=0$   characterizes the ductile response. Finally, when both conditions $g(x_0)=1$ and $g'(x_0)=0$ are met, we have critical behavior. While in the case of MS dynamics   all three of these possibilities can be realized,   in the brittle regime the GM response is characterized by a (Maxwell) jump taking place \emph{before} the spinodal point  \eqref{brittle} is reached. All these different possibilities  are illustrated in Fig.~\ref{fig:Asymptotics_Ductile}.

\subsection{Ductile behavior}
If the minimum of $h(x)$ is defined by the conditions   $g'(x_0)=0$ and $g(x_0)\neq 1$, we can   write,
\begin{equation}
h(x)\approx [1-g(x_0)]^2+g''(x_0)[1-g(x_0)](x-x_0)^2.
\end{equation}
Then the  application of the saddle-point approximation to  \eqref{eq:distr} gives  
\begin{equation}\label{eq:assympducspin}
p_a(\Delta)= \frac{1}{\sqrt{2\pi}} \, \Delta^{-3/2}[1-g(x_0)] p(x_0) \,e^{-\Delta [1-g(x_0)]^2}\sqrt{\frac{2 \pi}{\Delta \vert h''(x_0)\vert }}\propto \Delta^{-2} e^{-\Delta[1- g(x_0)]^2}.
\end{equation}

While  here the asymptotic behavior of the distribution is fully dominated by an exponential cut-off, the exponent value $-2$ can be still read of the empirical diagrams shown in Fig. \ref{fig:av} and Fig. \ref{fig:GM_Avalanches} for ductile behavior.

\subsection{Critical behavior}

The next special case is when  simultaneously $g(x_0)=1$ and $g'(x_0)=0$.  Since then  $h''(x_0)=0$, higher order terms have to be included into the expansion of the function $h(x)$. The third derivative term also vanishes, 
 $h'''(x_0)=0$,
and therefore  the Taylor expansion of   $h(x)$  starts with the fourth order term
$
h(x)\approx  (h^{(4)}(x_0)/4!)(x-x_0)^4,
$
where $h^{(4)}(x_0)=3 g''(x_0)^2.$ 
Moreover, we can write $ \phi(x)\approx(\phi''(x_0)/2)(x-x_0)^2,$  with $\phi''(x_0)=-p(x_0) g''(x_0)$, which   allows us to re-write the integral \eqref{eq:distr} as
\begin{equation}
p_a(\Delta)=   \frac{\Delta^{-3/2}}{\sqrt{2\pi}} \int_{x_i}^{x_{0}} -p(x_0)g''(x_0)(x-x_0)^2 e^{-\Delta\frac{3g''(x_0)^2}{4!}(x-x_0)^4} dx.
\end{equation}
Finally, performing the integration explicitly we obtain 
\begin{equation}
p_a(\Delta)= \frac{\Delta^{-3/2}}{\sqrt{2\pi}}\frac{p(x_0)g''(x_0)}{4 \left(\Delta\frac{3g''(x_0)^2}{4!} \right)^{3/4}}  \Gamma \left(\frac{3}{4},\Delta\frac{3g''(x_0)^2}{4!}(x-x_0)^4\right)\bigg\rvert_{x_i}^{x_{s}},
\end{equation}
where we introduced  the incomplete gamma function 
$
\Gamma(s,x)=\int_{x}^{\infty}t^{s-1}e^{-t}dt.
$
The lower limit of the integral does not contribute when  $\Delta$ is large and  we can write  the following asymptotic representation for  the avalanche size distribution
\begin{equation}
 p_a(\Delta) \sim  \Delta^{-\alpha},
\end{equation}
where $\alpha =9/4$. Note that  such critical regimes in the case of MS and GM dynamics are characterized by the same exponent   and therefore belong to the same universality class. Similar  'super universality'  for the random-field Ising model  was shown in \citep{PhysRevB.89.104201}. 

\subsection{Brittle behavior }

\paragraph{MS dynamics}
Suppose now that  $x_0$ solves  the equation  $g(x_0)=1$ which is precisely  Eq.~\eqref{eq:mxz}, identifying  the brittle regime. The avalanches must be counted till  the point  $x_0$ where the system undergoes a system size  macroscopic  failure.  We can  then  expand the function  $h(x)$ to obtain $ h(x)\approx \frac{h''(x_0)}{2}(x-x_0)^2$, where  $h''(x_0)=[g'(x_0)]^2$.
Since $g(x_0)=1$ we also have $\displaystyle {\phi(x_0)=\left[1-g(x_0)\right]p(x_0)/g(x_0)}=0,$ so in addition to  $h(x)$  we can also expand $\phi(x)$ to  obtain
$
\phi(x)\approx -g'(x_0)p(x_0)(x-x_0).
$
Using these results we  can  approximate  the integral \eqref{eq:distr} by
\begin{equation}
p_a(\Delta) =\frac{\Delta^{-3/2}}{\sqrt{2\pi}} \int_{x_i}^{x_{0}} g'(x_0)p(x_0)(x_0-x) e^{-\Delta\frac{g'(x)^2}{2}(x-x_0)^2} dx.
\end{equation}
Since we are in the brittle regime and  the avalanches are counted  up to $x=x_0$ we can compute  the Gaussian integral explicitly
\begin{equation}
p_a(\Delta) = \frac{\Delta^{-5/2}}{\sqrt{2\pi}} \frac{g'(x_0)p(x_0)}{\left[g'(x_0)\right]^2} e^{-\Delta\frac{g'(x)^2}{2}(x-x_0)^2}\bigg\rvert_{x_i}^{x_{0}}.
\end{equation}
At  large $\Delta$ the lower limit gives a vanishing contribution, which allows us to  write  the final  asymptotic formula for  the avalanche size distribution in the form $p_a(\Delta)  \sim   \Delta^{-\alpha} $,
with   $\alpha=5/2$. As we have already mentioned before, this value of the exponent $\alpha$   already appeared in the studies of the classical FBM loaded in the soft device \citep{hansen} and was associated with spinodal criticality \citep{Alava_2006}. Our model is an extension of this earlier model in the sense that the soft device case can be obtained in the  limit   $\Lambda   \rightarrow 0$. We can then conclude that the scaling in the  brittle regime is also of the spinodal type. Note, however,  that such spinodal  regime is not critical but super-critical in view  of the presence  of the system-size spanning events.
 
\paragraph{GM dynamics}
While the analysis presented above of the brittle regime is  valid for the case of MS dynamics, it cannot be automatically extended to the case of GM dynamics.  Indeed,  recall that the  condition $g(x_0)=1$, which coincides with Eq.~\eqref{eq:mxz}, states that the averaged response curve $z(x)$ has a (local) maximum.  However, we know that  along the GM path the system size failure  takes place   before the point  $x_0$.  In particular,  in the GM instability point  $x_*<x_0$, where the corresponding Maxwell threshold is reached,  the system is still metastable.   Therefore the  counting of avalanches should be interrupted at the point $x_*$ and in the integral \eqref{eq:distr} we must put  the upper limit at $x_{s}=x_*$, see Fig.~\ref{fig:Asymptotics_Ductile}d. Moreover, in this case, the function $h(x)$ will attain its minimum in the boundary point $x_*$ which slightly modifies the standard saddle point argument \citep{Bruijn}.  Indeed,  suppose  that at such point  $h'(x_*)>0$. Assume further that $h(x)\rightarrow \infty$, as $x\rightarrow -\infty$ and that the integral $\int_{-\infty}^{x_*}  e^{-h(x)}dx$ converges. Then, in the limit  $N\rightarrow\infty$    we can write  
$
\int_{- \infty}^{x_{*}}e^{-N h(x)}dx  \rightarrow  e^{N h(x_{*})}/(Nh'(x_{*})).  
$
This allow us to re-write the integral \eqref{eq:distr} in the form 
\begin{equation}
p_a(\Delta) \approx \frac{\Delta^{-3/2}}{\sqrt{2\pi}} {\displaystyle \frac{e^{-\Delta h(x_*)}}{\Delta h'(x_*)} \phi(x_*)}\approx \frac{\phi(x_*)}{h'(x_*)} \frac{\Delta^{-5/2}}{\sqrt{2\pi}} e^{-\Delta(h(x_*))}  .
\end{equation}
The desired scaling  relation is then 
 \begin{equation} 
 p_a(\Delta) \sim    \Delta^{-\alpha}e^{-\Delta(1-h(x_*))},
 \end{equation}
with $\alpha = 5/2$. Note that  the power law scaling is  compromised at the large avalanche sizes by an exponential cut off.  

All these analytical results, including the exact values of the exponents,  fully support  our  numerical computations  presented in Fig.~\ref{fig:av}  and Fig.~\ref{fig:GM_Avalanches}. We have then analytically shown the existence of two scaling exponents characterizing the behavior of our augmented fiber bundle model. 
While the classical exponent 5/2 is recovered in the brittle regime, the transition from brittle to the ductile regime is characterized by the new exponent 9/4. Such crossover criticality must be tuned as in the case of classical thermodynamical \emph{critical} point. Instead, the super-criticality observed in the brittle regime is \emph{robust} and can be associated with the presence of a \emph{spinodal} point.

\begin{figure}[t]
\centering
\includegraphics[scale =1]{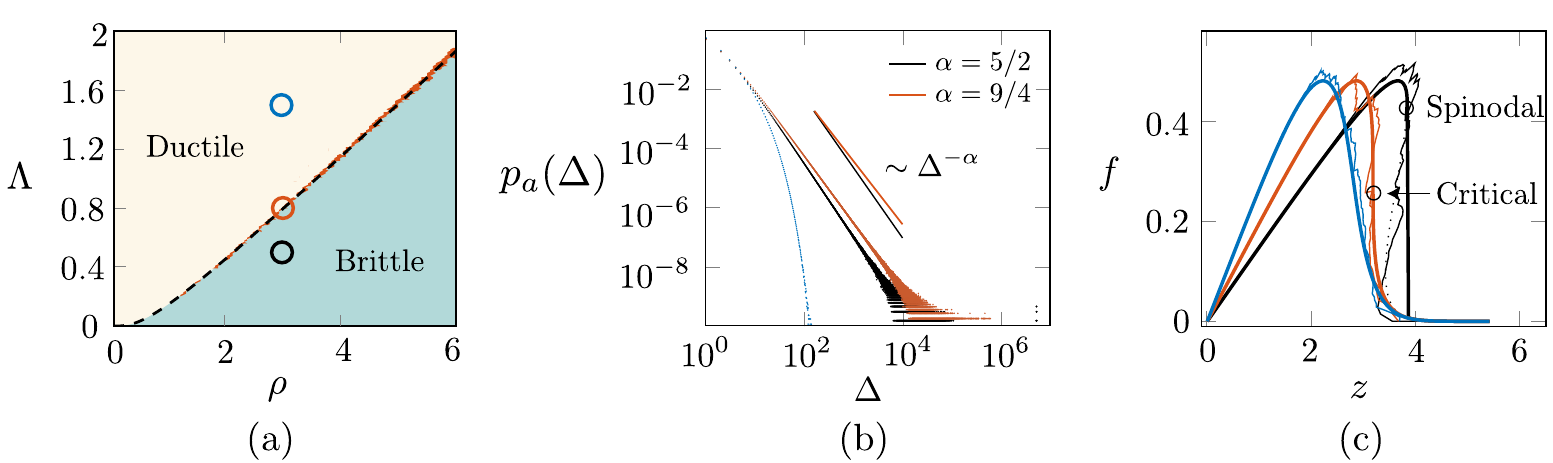}
 \caption{(a) Brittle-to-ductile transition (critical) line in the $(\rho, \Lambda)$ phase diagram. The colored circles in the phase diagram indicate the set of parameters used in (b) and (c). (b) The cumulative avalanche distribution for the the system with the corresponding parameters with $N=10^7$. (c) The superposition of the average force-elongation curve with a realization of system of size $N=100$. }
\label{fig:popsnap}
\end{figure}

In Fig.~\ref{fig:popsnap} we show   for the case of MS dynamics and Weibull distribution of thresholds the phase diagram in the  rigidity--disorder parameter space   exhibiting all three scaling regimes. The crossover region (red) was constructed by applying numerically the  Kolmogorov--Smirnoff criterion to the power law exponent which was found using the maximum likelihood method \cite{Clauset_2009, newman}.  More specifically, the analysis of the power-law was performed  in two steps. First, the parameter $\alpha$ was estimated using the maximum likelihood method. Second, we performed a goodness-of-fit test (using  the Kolmogorov method)  comparing  the obtained avalanche statistics  with  the power law ansatz found in the previous step.  This comparison produced  a so called $p$-value and we assumed that if  $p>0.1$, the power law hypothesis is  a plausible representation of the data. Finally,  the thick black line in Fig.~\ref{fig:popsnap}(a), tracing the crossover region, is obtained by solving the two equations  $g(x_0;\rho,\Lambda)=1$ and $g'(x_0;\rho,\Lambda)=0$, which is precisely Eq.~\eqref{eq:lfw} for the one-parameter Weibull distribution.

\begin{figure}[ht]
\centering
\includegraphics[angle=0, scale=.8]{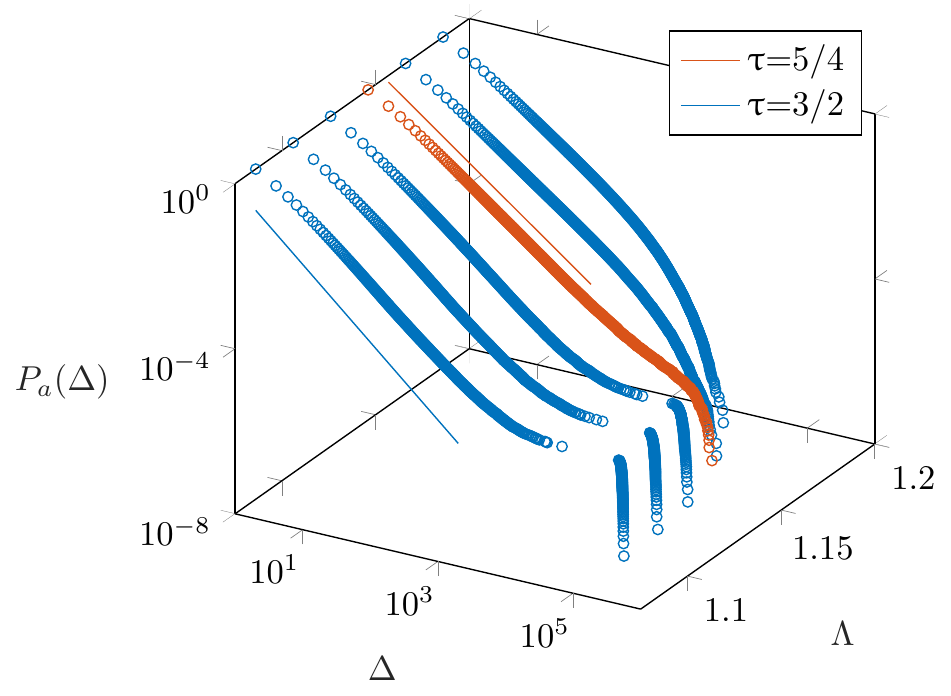}
\caption{Cumulative avalanche distribution showing the crossover from  ductile to brittle  regimes at  varying   $\Lambda$ and fixed  disorder   (under MS dynamic strategy)}\label{fig:3D Avalanches}
\end{figure}

  In  Fig.~\ref{fig:3D Avalanches} we use  the cumulative distribution of avalanches $P_a(x) = \int^x p_a(s)ds  \sim x^{-\tau}$, with  $\tau=\alpha-1$,   to illustrate  the crossover from the exponent  $\tau = 5/2-1=3/2$  associated with the spinodal criticality to the exponent $\tau = 9/4-1=5/4$ characterizing the  critical point.

\section{Finite size scaling}

So far, we have been mostly interested in the behavior of the system in the thermodynamic/continuum  limit. However,  in many applications of fracture mechanics there is a  need to understand the effects of finite system size. For instance, in cell cohesion/decohesion phenomena, the area density of proteins responsible for the binding/unbinding is often of the order of a few hundred   per $\mu m^2$ \citep{Bell_science_1978, Erdmann_ERJE_2007,Schwarz_RMP_2013, ES_PRL_2004}.  In such situations  it is of interest  to  study the effect on the statistics of avalanches of  the number of   elements $N$. When $N$  is finite, analytical approach fails and  we have to resort to numerical simulations.

\begin{figure}[ht]
\centering
\includegraphics[width = \textwidth]{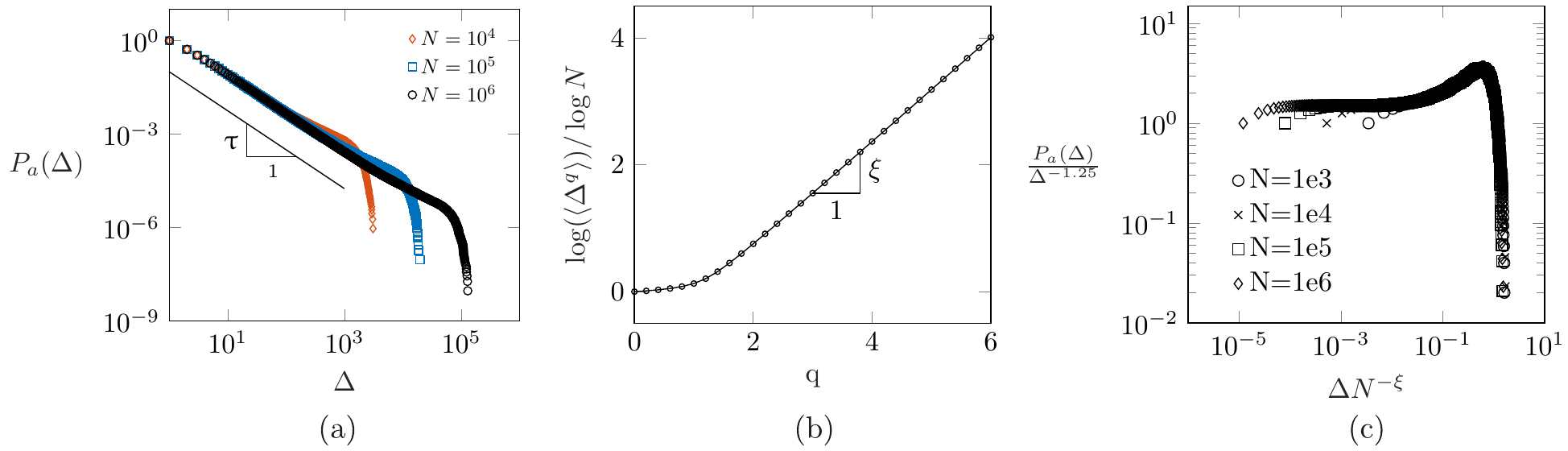}
\caption{(a) Avalanche cumulative distribution for the system in the critical state for several system sizes. (b) Plot of the moment  spectrum for the distribution of avalanches $\Delta$. The linear part has a slope $\xi\approx 0.8$. (c) Data collapse analysis for the avalanche distribution. The values for the critical exponents are $\tau=1.25$ and $\xi=0.8$. }
\label{fig:FSS}
\end{figure}

Yet, near the critical point some finite $N$ information can be obtained. To show this, suppose first that  the dimensionless  stiffness of the external spring $\lambda_f$, representing the elasticity of the effective environment, is  size independent.  In this case,  the size dependence of the critical  distribution of avalanches under  MS dynamic protocol  is illustrated  in Fig.~\ref{fig:FSS}. Clearly, the large event cut-off in this case is size dependent. 

To understand the structure of the cut-off functions, which are also  expected to be universal near the critical point,  we  write the \textit{cumulative }probability distribution in the form
\begin{equation}
P_a (\Delta)=\Delta^{-\tau} \mathcal{G}(\Delta/\Delta_c). 
\end{equation}
As the system size increases, the cutoff parameter  $\Delta_c$ is  expected to diverge  as $N^\xi$ with exponents $\left\lbrace \tau, \xi \right\rbrace$ both characterizing the universality class of the model. 

To test such finite size scaling (FSS) hypothesis and to find the critical exponent $\xi$, we performed numerical simulations starting with the values of parameters $(\rho,\Lambda)$   on the critical line   of the averaged  system. We conducted  simulations at several values of   $N$, adjusting  the values of  $(\rho,\Lambda)$ till  we   covered the whole range of system sizes of interest.  The analytically predicted  value of the cumulative   exponent   $\tau =  5/4$ was confirmed. The computation of the exponent $\xi$ was  performed through the standard method of moments  of $\mathbb{P}(\Delta)$ which reduces  to the checking of  the hypothesis that  
$\left< \Delta^q\right>   \sim N^{\xi(q+1-\tau)}$ \citep{chessa1999critical}.

In Fig.~\ref{fig:FSS}(b), we demonstrate that the conjectured FSS behavior emerges starting from $q=1.8$. This result is confirmed by   the data collapse for the re-scaled distribution $P_a(\Delta , N) \Delta^{\tau}$ vs. $\Delta N^{-\xi}$, where  we use the exponents $\tau=5/4$ and $\xi=0.8$  obtained from the moment  analysis, see Fig.~\ref{fig:FSS}(c).

Suppose now that  the dimensionless  stiffness  of the external spring $\lambda_f$ varies with the system size $N$.  Since  $ \lambda_{f}=  \kappa_{f} /\kappa N$,  this means that we no longer assume that $\kappa_f \sim N$.  Formally, the dimensional  stiffness $\kappa_f$ characterizes a single linear spring connected in series to the  parallel bundle of nonlinear breakable springs. However, the mean-field nature of the model hides the actual spatial structure of the system where each breakable element may interact indirectly with any other breakable element as it is the case, for instance, in a 2D RFM. Such interactions can be represented effectively in our model by various assumptions about the $N$ dependence of  $\kappa_f$. 

Assume that  $\kappa_f \sim  N^\beta$, where the exponent $0\leq \beta\leq 1$  can be seen as the (potentially  fractional) dimensionality of the load transmitting network.  The simplest assumption would be that there is indeed an external $N$-independent  elastic environment which would mean that $\beta=0$. Then $\Lambda = \lambda_f (\lambda+1)/\lambda\sim 1/N$.  Another limiting case, which we have studied in detail above,  is when  $\kappa_f \sim   N$  which would mean that  such external elasticity is extensive and $\beta=1$ so that  $\Lambda$ is size independent.  One can also argue that in the simple tension test for a 3D body whose volume scales as $ L^3\sim N $, the load is applied on a surface with dimension $L^2 \sim N^{2/3}$. Then,  if $\kappa_f$ is to represent the elasticity of the coupling between the load device and the body, we should have  $\kappa_f\sim  N^{2/3}$ which means that $\beta=2/3$ and $\Lambda\sim   N^{-2/3}$.

Consider, specifically, the case when   $\kappa_f$ is fixed and $\beta=0$. Then increasing $\Lambda$   at a fixed   $\lambda$    would mean decreasing $N$ and therefore increased ductility   can be viewed as a  small  size effect. We assume that  our analytical results targeting the  thermodynamic limit $N \to \infty$ can be still used to capture  such size effect if we simply set $\lambda_f \sim 1/N$. To show  that the  $N$ dependence of $\Lambda$ represents the main effect we   first write   $\Lambda = \tilde{\Lambda}/N$, and assume that the  effective rigidity of the system is now   characterized by the \emph{size independent} parameter $\tilde{\Lambda}$.  We can then construct the curves $z(x)$   at different $N$ but  with fixed  $\tilde{\Lambda}$. 
 
\begin{figure}[ht]
\centering
\begin{subfigure}[b]{0.45\textwidth}
\centering
\includegraphics[scale=0.9]{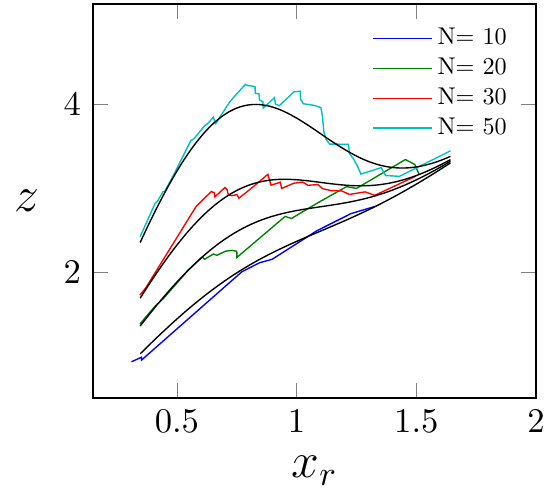}
\caption{ }\label{fig:zkkrho3kf10}
\end{subfigure}
\begin{subfigure}[b]{0.45\textwidth}
\centering
\includegraphics[scale=0.9]{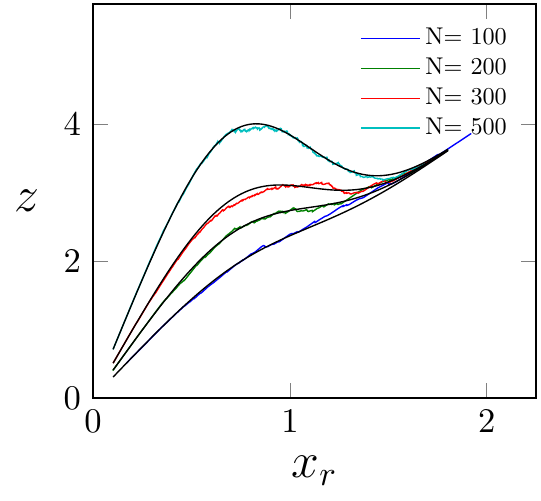}
\caption{}\label{fig:zkkrho3}
\end{subfigure}
\caption{The effect of the system size on the overall response of the system   for two values of the  fixed  rigidity
 measure: $\tilde{\Lambda}=20$(a) and  $\tilde{\Lambda}=200$ (b). The colored curves are single realizations of disorder;  in black we show the  averaged curves obtained by solving   \eqref{eq:zave}. 
}\label{fig:zave}
\end{figure}

In Fig.~\ref{fig:zave}(a, b) we compare the corresponding numerically obtained response curves for particular realizations of disorder  with the analytical results in the thermodynamic limit  where  we used  the parametric dependence on $N$ through  $\Lambda$. As we see,  the agreement is already good for $N \sim 100$, and particular  $\tilde{\Lambda}$. In this figure we see how the   BTD transition  is captured  as a size effect.  In contrast to what we have seen in the thermodynamic limit, now the BTD transition is not abrupt but is  represented by  an extended  \emph{critical region}.

%

\begin{figure}[ht]
\centering
\includegraphics[scale=1.0]{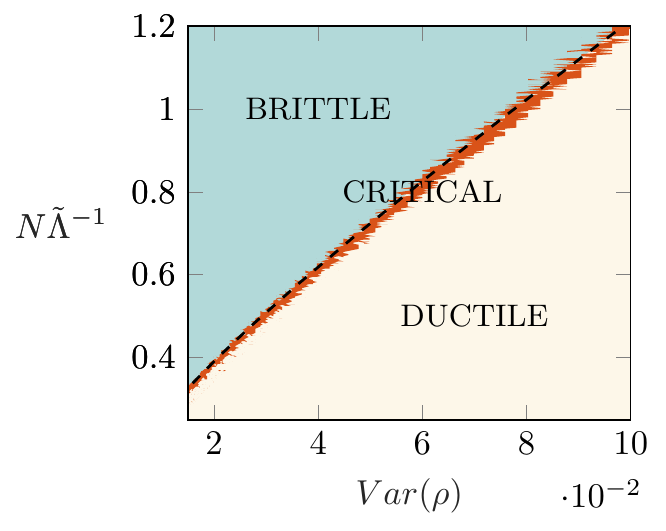}
\caption{The effect of the system size $N$ in the overall response of the bundle for  $\tilde{\Lambda}=2 \times 10^4$ and $\lambda=1$ as function of the variance of the disorder   $\text{Var} (\rho) = \Gamma\left(1+ 2/\rho\right)-\left[\Gamma\left( 1+ 1/\rho\right) \right]^2$ (Weibull distribution), see also a related empirical diagram in \cite{richard2021brittle}. The diffuse red region shows the uncertainty of interpreting statistics of avalanches as a power law.}
\label{fig:FSPD_VarX_Lambda}
\end{figure}

This region is clearly visible in Fig.~\ref{fig:FSPD_VarX_Lambda} where we show the phase diagram which  is based on direct numerical simulations of the system at a given $(\rho, N)$.  The diagram  shows three domains with a structurally different distribution of avalanches. In the  perforated (red) domain, we observed the power law distribution of avalanches with exponent $\alpha=9/4$ indicating criticality. In the brittle (blue)  domain, we could identify the super-critical avalanche distribution with exponent $\alpha=5/2.$   This is our regime of robust spinodal criticality.  Finally, in the ductile (yellow) domain, the power law scaling is  absent. To assess the quality of the power laws we estimated the scaling exponent using maximum likelihood method, and then tested the power law hypothesis using the Kolmogorov--Smirnov criterion \cite{Clauset_2009,newman,Vives_2012}.  
 
According to the phase diagram presented in Fig.~\ref{fig:FSPD_VarX_Lambda},   large systems will be brittle with pseudo-critical avalanche distribution dominated by a large catastrophic  (SNAP) events. This observation suggests that the ductile (quasi-brittle) regime  can only exist as   a finite size effect   disappearing in the thermodynamic limit, see also \cite{Dussi_PRL_2020}.   As the size  of the system decreases, the   spanning  events progressively disappear,  giving rise to  robust  scaling  with correlation length reaching the system size. Fluctuations (avalanches) take the form of a \emph{crackling} noise with a cut-off.  The real (percolation type) criticality in such setting can  be observed only in a single point corresponding to an infinite (brittle) system with an infinitely broad disorder.  As the system size is decreased further the correlation length becomes microscopic and the system enters the ductile (POP)  regime with largely Gaussian statistics of avalanches. Smaller systems naturally  place limits on the size of the bursts, which eliminates  large events and  extends the range of the post cut-off  ductile response.
  
Similar  size effect  was  identified  in numerical simulations of the  2D RFM  \citep{PhysRevLett.110.185505}, where  our ductile regime was interpreted as damage percolation, and the analog of our brittle regime was associated with crack nucleation;  in view of the implicit identification of the analog of our  rigidity  measure $\Lambda$ with $1/N$ the scaling in the crossover region  at a given disorder  was   interpreted as a finite size criticality.   The same type of   size dependence of the avalanche distribution was  also  observed in the study of self-organized criticality in a cellular automaton model of earthquakes  \citep{Olami};  other   related results can be found in  \citep{herrmann2014statistical, PhysRevE.73.046103, DELAPLACE199699, Herrmann_PRB_1989}.   In all considered cases, three types of breakdown processes were identified: localization, diffuse localization, and percolation-like regime. The numerical  study of transitions between different regimes  also suggested the existence of  an extended  region of intermediate system sizes where the critical scaling is   robust. This comparison shows that our analytically transparent   mean field model captures adequately all the main qualitative features of the more comprehensive   2D lattice models.

\section{Conclusions}

We used a  prototypical model of fracture in disordered solids to quantify the role of the system’s  rigidity
  (global connectivity) as a control parameter for the transition from brittle to ductile failure. The ductile response is usually associated with the stable development of small avalanches (micro-bursts), representing debonding events at the microscopic level. Instead, the brittle response is associated with large system-size events representing macro-crack-type system-size instabilities. In our model, the two types of fracture are distinguished by their statistical distribution of bond-breaking avalanches. 

More specifically, we showed that the avalanche statistics associated with brittle regimes could be interpreted as resulting from spinodal instability. These regimes, however, cannot be called critical because of the presence of system size events signifying global failure.  The complete criticality of crackling type \citep{SethnaDM01}, is observed exactly at the brittle-to-ductile transition,  which can be identified (at infinite disorder if $\beta=0$) with the classical thermodynamic critical point.  In the fully developed ductile regime, intermittency is lost and the distribution of fracture avalanches can be interpreted as a succession of uncorrelated events. 
 
One of our main conclusions is that the brittle-to-ductile transition can be associated with the crossover from spinodal to classical criticality, generating, in finite size systems, a scaling region with nonuniversal exponents. Such behavior is generic for a broad class of systems, encompassing fracture, plasticity, structural phase transitions, and even fluid turbulence. Our analysis reveals that in the transition region on the phase diagram, the system following the marginal stability (MS) dynamics exhibits an avalanche size distribution exponent different from the one associated with the robust spinodal criticality, characterizing brittle regimes. In the setting of global minimization (GM) dynamics, we observe a universal power-law distribution of avalanches only in the transition from brittle to the ductile regime with the same exponent as encountered under MS dynamics. One can then argue that the robust criticality, as in the case of earthquakes and collapse of compressed porous materials, should result from some feedback mechanisms ensuring self-tuning of the system towards the border separating brittle and ductile behaviors. Our study indicates that in addition to the strength of disorder, an appropriately chosen global measure of  rigidity can also serve as an instrument of such tuning.

The results obtained in our fully analytical study of the mean-field model of fracture in a disordered system can be used only for qualitative guidance in developing acoustic-emission-based precursors for global failure. Quantitative predictions should be, of course, based on the analysis of the comprehensive numerical models of realistic systems. And yet, we emphasize that the present prototypical model has already captured the principal features of such systems, including the presence of different scaling regimes in the space of disorder-rigidity parameters. Therefore, the obtained results can already serve as a basic guide in the design of new materials. In particular, our study unveils the possibility of using the statistical signature of stochastic fluctuations for non-destructive acoustic monitoring of the remaining life of fracturing samples.  By omitting some details,  available only through the study of purely numerical models,  this simple analytical model reveals the quantitative importance of collective effects and long-range correlations in fracture development. In this sense, it provides new fundamental insights into the mechanism of failure in disordered solids.

\section{Acknowledgments}
The authors are grateful to R. Garcia-Garcia and M. Mungan for helpful discussions.
H. B. R. was supported by a Ph.D. fellowship from Ecole Polytechnique; L. T. was supported by Grant No. ANR-10-
IDEX-0001-02 PSL.

\bibliographystyle{elsarticle-num}

\bibliography{bibfile}
\end{document}